\documentclass[epj, final]{svjour}

\usepackage{amsmath}
\usepackage{amssymb}
\usepackage{epsfig}

\newcommand{\mr}[1]{{\mathrm{#1}}}
\newcommand{\bfig}{\begin{figure}}
\newcommand{\efig}{\end{figure}}
\newcommand{\Tc}{T_{\mathrm{c}}}

\epsfclipon

\newcommand{\rsfig}[1]{\begin{center} 
                       \epsfig{file=#1, width=0.425\textwidth}
                       \end{center}
                       }

\begin{document}

\title{Glassy Dynamics of Simulated Polymer Melts:\\
Coherent Scattering and Van Hove Correlation Functions}
\subtitle{Part I: Dynamics in the $\beta$-Relaxation Regime}

\titlerunning{$\beta$-Dynamics of Glassy Polymer Melts}

\author{M.\ Aichele\inst{1} \fnmsep \thanks{Corresponding author. E-mail: \textsf{Martin.Aichele@uni-mainz.de}} \and J.\ Baschnagel\inst{2}
} 

%
%

\institute{Institut f{\"u}r Physik, Johannes Gutenberg-Universit{\"a}t Mainz, Staudinger Weg 7, 55099 Mainz, Germany \and Institut Charles Sadron, 6 rue Boussingault, 67083 Strasbourg Cedex, France}
\date{Received: date / Revised version: date}
%

\abstract{
We report results of molecular-dynamics simulations of a model polymer melt consisting of short non-entangled chains in the supercooled state above the critical temperature $\Tc$ of mode-coupling theory (MCT).  To analyse the dynamics of the system we computed the incoherent, the collective chain and the collective melt intermediate scattering functions as well as their space Fourier transforms, the van Hove correlation functions.  In this first part of the paper we focus on the dynamics in the $\beta$-relaxation regime.  The final structural relaxation, the $\alpha$-relaxation, will be studied in the following second part.  The results can be summarized as follows: Without using any fit procedure we find evidence for the space-time factorization theorem of MCT in real and reciprocal space, and also for polymer specific quantities, the Rouse modes.  The critical amplitudes in real space are determined directly from the simulation data of the van Hove correlation functions.  They allow to identify the typical length scales of the $\beta$-dynamics, and illustrate that it is a localized process.  In a quantitative analysis the wave-vector dependences of the $\beta$-coefficients, i.e., of the non-ergodicity parameter, the critical amplitude, and the next-to-leading order correction coefficients, are studied for all correlators.  The $\beta$-coefficients show indications of polymer specific effects on the length scale of the chain's radius of gyration.  The agreement between simulation and the leading-order MCT description is found to be good in the central $\beta$-regime.  Next-to-leading order corrections extend the validity of the MCT approximations to a greater time window and become more important at large wave-vectors.  
\PACS{
      {64.70.Pf}{Glass transitions}   \and
      {61.25.Hq}{Macromolecular and polymer solutions; polymer melts; swelling} \and
      {61.20.Ja}{Computer simulation of liquid structure}
     } 
} 
\maketitle


%
\section{Introduction}
\label{sec:intro}
A distinguishing property of glass forming liquids is the strong increase of the structural relaxation time, which precedes the glass transition, i.e., the solidification of the liquid on the experimental time scale at the glass transition temperature $T_\mr{g}$ \cite{Vigo1997,Pisa1998,Trieste1999}.  Understanding the microscopic origin of this remarkable slowing down of the dynamics represents one of the most challenging problems in condensed matter physics.  A solution to this problem was proposed by mode-coupling theory (MCT) \cite{Goetze1999_review,goetzemctessentials,GoetzeSjoegren1995_TTSP,Goetze_LesHouches}.  The theory predicts the existence of a critical temperature $\Tc$, well above $T_\mr{g}$, at which the dynamics qualitatively changes.  Above, but close to $\Tc$ the dynamics is dominated by the mutual blocking of a particle and its neighbours (``cage effect'').  This blocking creates (free) energy barriers which would become infinite and lead to a cease of any structural relaxation if the cage effect determined the dynamics alone (idealized MCT).  However, very close to and below $\Tc$ additional relaxation mechanisms gradually become more important than the cage effect.  The approximate inclusion of these processes in the theory (extended MCT) avoids the absolute freezing at $\Tc$ \cite{GoetzeSjoegren1995_TTSP,FuchsGoetzeHildebrand1992_extMCT}.  The glass former remains liquid down to $T_\mr{g}$, where it falls out of equilibrium.  According to MCT, $\Tc$ is a characteristic (thermodynamic) temperature of glass forming liquids.  

The existence of an additional characteristic temperature is suggested by recent attempts to connect the structural glass transition to mean-field spin-glass models with discontinuous order parameter (see \cite{KirkpatrickThirumalai_TTSP1995,MezardParisi2000} for a review). This approach is motivated by two observations: First, the long-time dynamics of these models is described by the idealized mode-coupling equations of schematic models.  Therefore, they exhibit a dynamic transition at a temperature which is identified with $\Tc$.  Second, they also undergo a static transition to a spin-glass phase at a temperature $T_\mr{s}$ below $\Tc$.  In the interval $T_\mr{s} < T < \Tc$ the dynamics is determined by an exponentially large number of local free energy minima.  The system stays in one minimum for a long time before escaping to another one by activated processes.  Below $T_\mr{s}$ the number of minima becomes finite and the system is trapped in one of them.

At present, it is not clear whether this analogy between spin and structural glasses complements or challenges mode-coupling theory.  A recent application to experiments \cite{KrakoAlba2000} shows that the theory overestimates $T_\mr{s}$. It is placed in the liquid phase close to $\Tc$.  On the other hand, this application as well as other experimental or simulation studies \cite{Vigo1997,Pisa1998,Trieste1999,Goetze1999_review} provide evidence for the existence of a critical temperature $\Tc$, as predicted by MCT.  Such evidence has motivated extensions of MCT to treat orientational degrees of freedom \cite{SchillingScheidsteger1997,FabbianLatz2000,FranoschGoetze_orient1997}, vibrations \cite{FranoschGoetze1997,goetzevoigtmann2000} or non-equilibrium systems \cite{Latz2000}.  Furthermore, many computer simulations \cite{kobreview1999} have been undertaken to understand better the dynamics above and below $\Tc$ (dynamic heterogeneity \cite{DonatiGlotzer1999,Allegrini1999,YamamotoOnuki1998,DoliwaHeuer1998,DoliwaHeuer2000,VollmayrKob2000}, connection between the potential energy landscape and structural relaxation \cite{BuechnerHeuer_PRE1999,BuechnerHeuer_PRL2000,SastryDebenedetti1999,SchroderSastry2000,SciortinoKobTartaglia2000}, physical aging \cite{KobBarrat2000,KobSciortinoTartaglia2000}), but also to further test the theory on other systems than simple liquids. Some examples are diatomic molecules \cite{KaemmererKobSchilling1998,KaemmererKobSchilling1998_orient,KaemmererKobSchilling1997,TheisSciortino2000}, ortho-terphenyl \cite{LewisWahnstrom1994,Mossa2000}, water \cite{sciortino1996,SciortinoFabbianChen1997,Sciortino2000,StarrSciortino1999} or polymers \cite{ZonLeeuw1998,ZonLeeuw1999,BBPB_2000}.   

With this and the subsequent paper \cite{alphaDynamics} (hereafter called part I and part II, respectively) we want to contribute to this research by extending our previous analysis of a simple model for a supercooled polymer melt.  This analysis considered the relaxation behavior of the incoherent intermediate scattering function \cite{BennemannBaschnagelPaul1999_incoherent}, the dynamics of the model under isobaric and isochoric conditions \cite{BennemannPaulBinder1998,BennemannPaulBaschnagel1999}, the interplay between the cage and polymer-specific effects \cite{BBPB_2000,BennemannPaulBaschnagel1999_Rouse}, and correlated motion of the monomers close to $\Tc$ \cite{natureBDBG1999,Aichele_DH2000}.  The present work attempts to complement these studies by discussing results for the intermediate coherent scattering and van Hove correlation functions of both the polymer and the melt.  Whereas part II deals with the final structural relaxation of the $\alpha$-regime, the present part I describes the dynamics of the model in the $\beta$-relaxation regime defined by MCT close to $\Tc$.  It is organized as follows: Section~\ref{sec:model} briefly introduces the model.  Section~\ref{sec:quantities+theory} compiles the quantities studied and the theoretical background for the analysis. The following section~\ref{sec:results} summarizes the results and the final section~\ref{sec:conc} contains our conclusions.

\section{Model}
\label{sec:model}
In this section we briefly describe the model underlying our simulation.  A more detailed description can be found in \cite{BennemannPaulBinder1998}.  

We investigated a bead-spring model of linear polymer chains.  All monomers interact via a truncated and shifted Lennard-Jones (LJ) potential given by 
\begin{equation}
U_\mr{LJ}(r) := \left\{ \begin{array}{r@{,\quad}l}
4\epsilon  \left[ \left(\frac{\sigma}{r}\right)^{12}
- \left(\frac{\sigma}{r}\right)^{6}\right] + C & r < 2 r_\mr{min}\\
0 & r \geq 2 r_\mr{min}
\end{array} \right. \;,
\end{equation}
where the constant $C=0.00775$ is chosen such that the potential vanishes continuously at $r = 2 r_\mr{min}$,  $r_\mr{min} = 2^{1/6}\sigma$ $= 1.1225 \sigma$ being the minimum of $U_\mr{LJ}$.  Throughout this paper all quantities are measured in Lennard-Jones units: temperature and distances were measured in units of $\epsilon /k_\mr{B}$ and $\sigma$, respectively, and time in units of ${(m \sigma^2 \epsilon)}^{1/2}$, with mass $m$ set to unity. 

Along the polymer backbone an additional FENE (finitely extensible nonlinear elastic) potential \cite{KremerGrest1990} was used to introduce bonds between adjacent monomers,
\begin{equation}
U_\mr{FENE}(r)=-\frac{k}{2}R_{0}^2 \ln \left[1-\left(\frac{r}{R_0}\right)^2 \right]
\end{equation}
with $R_0=1.5$ and $k=30$.  The superposition of the LJ and FENE potentials leads to an effective bond potential with a sharp minimum at $r_\mr{bond} = 0.9606$, making bond crossings impossible.  

These model parameters yield two incompatible length scales: $r_\mr{bond} = 0.9609$ and $r_\mr{min} = 1.1225$.  ``Incompatible'' means that it is not possible to arrange the beads in a regular fcc (or bcc) lattice structure if one requires that the beads take precisely these distances which yield the pairwise minima of the potentials.  Another feature of the model is that the chains do not become stiffer with decreasing temperature.  This is reflected by the nearly constant end-to-end distance $R_\mr{e}^2 = 12.3 \pm 0.1$ and radius of gyration $R_\mr{g}^2 = 2.09 \pm 0.01$ in the investigated temperature range \cite{BennemannBaschnagelPaul1999_incoherent}.  The flexibility along the backbone of the chain and the incompatibility of $r_\mr{bond}$ and $r_\mr{min}$ prevent crystallization of the melt \cite{MeyerMuellerPlate2001}.  The static structure exhibits the typical features of an amorphous material in the temperature range studied (see Fig.~\ref{fig:static_structure_factor_melt+chain+Debye}).  This is an essential premise for investigating glassy dynamics.
 
The simulations were done in two steps: First, the volume of the simulation box was determined in a constant pressure simulation at $p=1$.  Then, this volume was kept fixed and the simulations were continued in the canonical ensemble using the Nos{\'e}-Hoover thermostat.  Periodic boundary conditions were applied in all three spatial directions of the cubic simulation box.  A box typically contained between 110 and 120 chains with $N=10$ monomers each.  We simulated at $T=0.46$, 0.47, 0.48, 0.50, 0.52, 0.55, 0.6, 0.65, and 0.7.  While the lowest simulated temperature $T=0.46$ is slightly above $\Tc \simeq 0.45$, the critical temperature of the mode coupling theory \cite{Goetze1999_review,goetzemctessentials,GoetzeSjoegren1995_TTSP,Goetze_LesHouches}, the highest temperature $T=0.7$ leaves the temperature range where the $\beta$-dynamics can be observed \cite{BennemannBaschnagelPaul1999_incoherent}.

\section{ Simulated Quantities and Theoretical Background}
\label{sec:quantities+theory}
The static structure of a glass is almost indistinguishable from that of a liquid.  The two phases mainly differ in the values of their relaxation times: The relaxation time in the liquid phase is small, but becomes unmeasurably large in the glassy phase.  A means to investigate this crossover from fast to strongly protracted structural relaxation are dynamic correlation functions.  In this paper we present our findings for the coherent intermediate scattering and the van Hove correlation functions in the $\beta$-relaxation regime.  In the following subsection \ref{subsec:quantities} these functions are defined and their calculation from simulation data is briefly described.

In the vicinity of $\Tc$ mode-coupling theory (MCT) predicts a two-step relaxation process for correlation functions which couple to density fluctuations \cite{Goetze1999_review,goetzemctessentials,GoetzeSjoegren1995_TTSP,Goetze_LesHouches}.  Some MCT results are compiled in the second subsection \ref{subsec:theory} as well as approximations for the single chain scattering function, which are motivated by polymer theory \cite{DoiEdwards}.

\subsection{Analysed Quantities}
\label{subsec:quantities}

The coherent intermediate scattering function of the melt is defined by
\begin{equation}
\label{eq:def_phiqt}
\phi_q(t) := \frac{S_q(t)}{S_q(0)} \;.
\end{equation}
Here, $S_q(0) \equiv S_q$ is the collective static structure factor and $S_q(t)$ is given by
\begin{equation}
\label{eq:defsqt}
S_q(t) := \frac{1}{M} \Bigl\langle \sum_{i=1}^M\sum_{j=1}^M \exp \left\{ \mr{i} \vec{q}\cdot\bigl[\vec{r}_i(t)-\vec{r}_j(0)\bigl]\right\} \Bigl\rangle \;,
\end{equation} 
where $\vec{r}_i(t)$ is the position of the $i$th monomer at time $t$ in the melt, which contains $M$ monomers in total, and $\langle \cdot \rangle$ denotes the canonical ensemble average. 

Due to isotropy all discussed quantities depend on the modulus $q = |\vec{q}|$ of the wave vector $\vec{q}$ only.  In the following we will use a superscript ``s'' for incoherent (self) and a superscript ``p'' for chain (polymer) quantities.  Coherent quantities of the melt do not carry a superscript.   $\phi^\mr{x}_q(t)$ stands for any scattering function, where the superscript ``x'' indicates the different correlators.  In the case of the chain scattering function, $\phi^{\mr{p}}_q(t)$ is defined by equations~(\ref{eq:def_phiqt}) and (\ref{eq:defsqt}) when using $N$, the number of monomers per polymer, instead of $M$ and summing over all pairs of monomers belonging to the same chain.  Setting $i=j$ in equation~(\ref{eq:defsqt}) defines $\phi^\mr{s}_q(t)$, the incoherent scattering function.

Since the system is homogeneous and isotropic, one can simplify the formulas in a way that is more suitable for computation.  For instance, $S_q(t)$ may be calculated by
\begin{equation}
\begin{split}
S_q(t) = &\frac{1}{M} \Bigl\langle \sum_{i=1}^M{\cos[\vec{q}\cdot\vec{r}_i(t)]} \sum_{j=1}^M{\cos[\vec{q}\cdot\vec{r}_j(0)]}\\
 &+ \sum_{i=1}^M{\sin[\vec{q}\cdot\vec{r}_i(t)]}\sum_{j=1}^M{\sin[\vec{q}\cdot\vec{r}_j(0)]} \Bigl\rangle \;,
\end{split}
\end{equation}
which allows computation with a number of trigonometric operations of order $M$, instead of order $M^2$.  For $S^\mr{p}_q(t)$ and $S^\mr{s}_q(t)$ analogous expressions are obtained.  

The set of $q$-values for which the scattering functions were calculated is $q \in \{1.0$, 2.0, 3.0, 4.0, 5.0, 6.0, 6.9, 7.15, 8.0, 9.5, 11.0, 12.8, 14.0, 16.0, $19.0\}$.
These values cover the maxima and minima of the static structure factor (cf.\ Figure~\ref{fig:static_structure_factor_melt+chain+Debye}).  Especially for small $q$ there are only very few reciprocal vectors complying with the periodic boundary conditions.  This leads to unsatisfactory statistics in the case of the scattering function of the melt (for $S^\mr{p}_q(t)$ the statistics is $M/N$ times better).  In order to improve the statistics the scattering functions were averaged over an interval of width $\Delta$ around $q$. So, whenever quoting a function value $\phi^\mr{x}_q(t)$, this actually means
\begin{equation}
\label{eq:binning}
\phi^\mr{x}_q(t) \approx \frac{\sum_{\{\vec{q} \,| q \in [q-\Delta/2,\, q+\Delta/2]\}} \phi^\mr{x}_{\vec{q}}(t)}{\bigl|\bigl\{\vec{q} \,| q \in [q-\Delta/2, q+\Delta/2]\bigr\}\bigr|}\;,
\end{equation}
where the sum runs over all allowed reciprocal vectors.  $\Delta$ was chosen such that it is as small as possible, but provides sufficient statistics, and it was checked that the averaging did not introduce significant deviations. 

Insight in the dynamics in real space can be obtained by analysing the van Hove correlation functions, which are the Fourier transforms of the respective scattering functions.  For the melt the van Hove correlator is defined by (see e.g.\ \cite{HansenMcDonald})
\begin{equation}
G(\vec{r},t) := \frac{1}{M} \Bigl\langle \sum_{i=1}^M\sum_{j=1}^M{\delta[\vec{r} + \vec{r}_j(0) - \vec{r}_i(t)]} \Bigl\rangle \;.
\end{equation}
$G(\vec{r},t)$ is proportional to the probability density for finding a particle in the volume element $\mr{d}\vec{r}$.  It is useful to separate $G(\vec{r},t)$ into self and distinct parts, $G(\vec{r},t) = G_{\mr{s}}(\vec{r},t) + G_{\mr{d}}(\vec{r},t)$
with
\begin{align}
\label{eq:defvanhoveself}
G_\mr{s}(\vec{r},t) &= \frac{1}{M} \Bigl\langle \sum_{i=1}^M{\delta[\vec{r} + \vec{r}_i(0) - \vec{r}_i(t)]} \Bigl\rangle \; , \\
\label{eq:defvanhovedist}
G_\mr{d}(\vec{r},t) &= \frac{1}{M} \Bigl\langle \sum_{i=1}^M \sum_{\substack{j=1\\j\neq i}}^M \delta[\vec{r} + \vec{r}_j(0) - \vec{r}_i(t)] \Bigl\rangle \;. 
\end{align}
Hence, we have at $t=0$
\begin{equation}
 G_{\mathrm{s}}(\vec{r},0) = \delta(\vec{r}), \quad G_{\mathrm{d}}(\vec{r},0) = \frac{M-1}{V} g(\vec{r}) \;,
\end{equation}
where we introduced the pair correlation function $g(\vec{r})$
\begin{equation}
g(\vec{r}) := \frac{V}{M (M-1)} \Bigl\langle \sum_{i=1}^M \sum_{\substack{j=1\\j\neq i}}^M \delta[\vec{r} + \vec{r}_j - \vec{r}_i] \Bigl\rangle \;.
\end{equation}
The denominator $M(M-1)$ is the total number of monomer pairs in the melt.  For polymers all quantities are defined in the same way, with $M$ replaced by $N$, and indicated by the superscript ``p''.  Again, we introduce a short hand notation, $G_\mr{X}(\vec{r},t)$, to denote self and distinct van Hove functions.  In the case of isotropic systems $G(\vec{r},t)$ depends on $r = |\vec{r}|$ only.  After discretization the van Hove functions can be computed from histograms of pair distances in the melt, taking periodic boundary conditions into account.

\subsection{Theoretical Background}
\label{subsec:theory}
This section is split into two parts.  The first part \ref{subsubsec:mct} summarizes those MCT predictions which are relevant for the subsequent analysis of the scattering and van Hove correlation functions, whereas the second part \ref{subsubsec:rouse} discusses approximations, suggested by the theory of polymer dynamics \cite{DoiEdwards}, for the coherent scattering function of a chain.

\subsubsection{Mode-Coupling Theory}
\label{subsubsec:mct}

The subsequent discussion mainly follows references \cite{FranoschFuchsGoetze1997,FuchsGoetzeMayr1998}, where one can also find quantitative results for a system of ideal hard spheres.

The central prediction of MCT is the existence of a critical temperature $\Tc$ in the temperature regime of the supercooled liquid.  In the vicinity of $\Tc$ asymptotic results can be derived for small values of the separation parameter $\sigma$,
\begin{equation}
\sigma := C \,\frac{\Tc-T}{\Tc} \;.
\end{equation}
The constant $C$ is system dependent and typically of order 1 \cite{FranoschFuchsGoetze1997}.  The following equations are predictions in leading order of $|\sigma|$. 

In the idealized MCT the separation parameter determines the temperature dependence of the two relevant time scales of the theory, i.e., of the $\beta$-relaxation time $t_\sigma$,
\begin{equation}
\label{eq:beta_time_law}
t_{\sigma} := \frac{t_0}{|\sigma|^{1/2a}} \;,
\end{equation}
and of the universal $\alpha$-timescale $\tilde{\tau}$,
\begin{equation}
\label{eq:alpha_time_law}
\tilde{\tau} := \frac{t_0}{|\sigma|^{\gamma}} \;, \quad T \geq \Tc \;.
\end{equation}
In equations~(\ref{eq:beta_time_law}) and (\ref{eq:alpha_time_law}) $t_0$ represents a matching-time to the microscopic transient, and the exponents $a$ ($0 < a < 1/2$) and $\gamma$ are related to each other by $\gamma = 1/(2a) + 1/(2b)$.  Here, $b$ denotes the von Schweidler exponent which is in turn connected to the critical exponent $a$ by the exponent parameter $\lambda$ via
\begin{equation}
\lambda = \frac{\Gamma(1-a)^2}{\Gamma(1-2a)} = \frac{\Gamma(1+b)^2}{\Gamma(1+2b)}\;.
\end{equation}
The exponent parameter, and thus also $a$, $b$, and $\gamma$, depend on the glass former under consideration, but not on temperature or on the correlator $\phi^\mr{x}_q(t)$ studied. 

Close to $\Tc$, $\phi^\mr{x}_q(t)$ is predicted to decay in two steps: A first step leads to a plateau value, the so-called non-ergodicity parameter $f^\mr{x c}_q$, and a second step leads off of it.  The second step represents the initial part of the final structural $\alpha$-relaxation, whose temperature dependence is determined by $\tilde{\tau}$.  In this scenario the $\beta$-relaxation regime corresponds to the time-window between $t_0$ and $\tilde{\tau}$ ($t_0 \ll t \ll \tilde{\tau}$), and so to times on the scale $t_{\sigma}$, where $|\phi^\mr{x}_q(t) - f^\mr{x c}_q| \ll 1$.

The time-evolution of $\phi^\mr{x}_q(t)$ in the $\beta$-regime can be written as \cite{FranoschFuchsGoetze1997}
\begin{equation}
\label{eq:quantbetastart}
\begin{split}
\phi^\mr{x}_q(t) = &f^\mr{x c}_q + \frac{h^\mr{x}_q t_0^{\,a}}{t_\sigma^{\,a}}g(\hat{t})\\
 &+ \frac{h^\mr{x}_q t_0^{\,a}}{t_\sigma^{\,2a}}\left\{t_0^{\,a} B^2 \left[\kappa(-b) + K^\mr{x}_q\right]\right\}\hat{t}^{\,2b}\\
&+ \frac{h^\mr{x}_q t_0^{\,a}}{t_\sigma^{\,2a}}\left\{t_0^{\,a}\left[\kappa(a) + K^\mr{x}_q\right]\right\}\hat{t}^{-2a} \;,
\end{split}
\end{equation}
where $\hat{t} := {t}/{t_{\sigma}}$.  The first two terms of equation~(\ref{eq:quantbetastart}) constitute the so-called ``factorization theorem''.  The name derives from the property that the function $g(\hat{t})$ carries the whole temperature and time dependences, while the dependence on space and on the correlator enters only via $f^\mr{x c}_q$ and the critical amplitude $h^\mr{x}_q$.  Therefore, $g(\hat{t})$ is called $\beta$ master function.  Its shape is solely determined by the exponent parameter $\lambda$.  

The factorization theorem is the MCT result in leading order of $|\sigma|$ (it is of order ${|\sigma|}^{1/2}$).  The second and third lines of equation~(\ref{eq:quantbetastart}) represent corrections (of order $|\sigma|$) to it.  The third term of the sum (\ref{eq:quantbetastart}) is the next-to-leading order long-time correction, which becomes important in the late $\beta$-regime, whereas the last term gives the corresponding correction for the early $\beta$-relaxation.  These corrections violate the factorization property of the leading order result due the $q$-dependence of $K^\mr{x}_q$.  Expressions for $K_q$, $K^\mr{s}_q$ and $\kappa$ have been derived in \cite{FranoschFuchsGoetze1997,FuchsGoetzeMayr1998}.  For the analysis of the simulation data, it is only important that these quantities are given in terms of the static structure at $\Tc$ and can thus be considered as independent of temperature (in the same way as $f^\mr{x c}_q$ and $h^\mr{x}_q$).

To perform this analysis we rewrite equation~(\ref{eq:quantbetastart}) by combining those theoretical quantities which cannot be determined separately in a fit procedure.  Since the microscopic time $t_0$ is unknown, only the product
\begin{equation}
\label{eq:def_h_fit}
h^\mr{x \; fit}_q := h^\mr{x}_q t_0^{\,a}
\end{equation}
can be fitted.  Furthermore, we introduce long- and short-time correction coefficients, $A^\mr{x}_q$ and $B^\mr{x}_q$, defined by
\begin{align}
\label{eq:def_A_A}
A^\mr{x}_q &:= t_0^{\,a} B^2 \left[\kappa(-b) + K^\mr{x}_q\right]\;,\\
\label{eq:def_B_A}
B^\mr{x}_q &:= t_0^{\,a}\left[\kappa(a) + K^\mr{x}_q\right] \;. 
\end{align}
This yields the following expression for equation~(\ref{eq:quantbetastart})
\begin{equation}
\label{eq:quantbetaformula}
\phi^\mr{x}_q(t) = f^\mr{x c}_q + \frac{h^\mr{x \; fit}_q}{t_\sigma^{\,a}}g(\hat{t}) + \frac{h^\mr{x \; fit}_q}{t_\sigma^{\,2a}} A^\mr{x}_q \hat{t}^{\,2b} + \frac{h^\mr{x \; fit}_q}{t_\sigma^{\,2a}} B^\mr{x}_q \hat{t}^{-2a} \;.
\end{equation}
In this equation only $t_\sigma$ varies with temperature according to equation~(\ref{eq:beta_time_law}). All other quantities are (asymptotically) independent of temperature.  Futhermore, note that $A^\mr{x}_q$ and $B^\mr{x}_q$ satisfy the following linear relationship
\begin{equation}
\label{eq:B_A+A_A}
B^\mr{x}_q = \frac{A^\mr{x}_q}{B^2} + [\kappa(a) - \kappa(-b)]t_0^{\,a} \;.
\end{equation}
Both correction coefficients, $A^\mr{x}_q$ and $B^\mr{x}_q$, should thus exhibit the same $q$-dependence.  Equation~(\ref{eq:B_A+A_A}) can either be used as a test of this prediction or for calculating the short time coefficient $B^\mr{x}_q$ from the long-time coefficients $A^\mr{x}_q$, which were easier to obtain from the simulation data (see section~\ref{subsubsec:quantbeta}).

From a previous analysis of the incoherent scattering function \cite{BennemannBaschnagelPaul1999_incoherent} the exponents $a$, $b$ and $\gamma$, the $\beta$-relaxation times $t_{\sigma}$, the parameter $B$, and the $\beta$ master function $g(\hat{t})$ are known.  In section \ref{subsubsec:quantbeta} we want to exploit this knowledge as much as possible to extend the quantitative description of the $\beta$-dynamics of $\phi^\mr{s}_q(t)$ to coherent scattering.

\subsubsection{Coherent Chain Scattering Function and Rouse Model}
\label{subsubsec:rouse}

In a dense polymer melt excluded volume and hydrodynamic interactions are screened \cite{DoiEdwards}.  If such long-range interactions are absent, the Rouse model is commonly assumed to provide a viable approximation for the dynamics of non-entangled chains \cite{KremerGrest_review1995,BinderPaul_review1997}.  Although the entanglement length $N_\mr{e}$ has not been determined for the present model yet, extensive simulation studies of a closely related model suggest $N_\mr{e} \approx 32$ \cite{KremerGrest1990,PuetzKremerGrest2000}.  Since our simulations are done with $N=10$ ($< N_\mr{e}$), the Rouse model should apply.  Therefore, we want to compare the coherent chain scattering function with the predictions of the Rouse theory in the following.

A basic assumption of the Rouse model is that every monomer experiences a local random force which is Gaussian distributed.  Since the displacement $\vec{r}_i(t) - \vec{r}_j(0)$ is a linear function of these random forces, it is also a Gaussian random variable \cite{DoiEdwards}. This implies that the canonical average of the exponential in equation~(\ref{eq:defsqt}) can be written as
\begin{multline}
\left\langle \exp\left\{\mr{i}\vec{q} \cdot \bigl[ \vec{r}_i(t) - \vec{r}_j(0)\bigr] \right\} \right\rangle \\
= \exp\left\{ - \frac{q^2}{6} \left\langle \bigl[ \vec{r}_i(t) - \vec{r}_j(0)\bigr]^2 \right\rangle \right\} \;
\end{multline}
so that
\begin{equation}
\label{eq:gauss_approx}
S^\mr{p}_q(t) = \frac{1}{N} \sum_{i=1}^N\sum_{j=1}^N \exp\left\{ - \frac{q^2}{6} \left\langle \bigl[ \vec{r}_i(t) - \vec{r}_j(0)\bigr]^2 \right\rangle \right\} \;.
\end{equation}
Note that the time-dependence of this ``Gaussian approximation'' for $S^\mr{p}_q(t)$ is completely given in terms of the mean-square displacements $\langle [\vec{r}_i(t) - \vec{r}_j(0)]^2 \rangle$ and that the $q$-dependence is quite different from that of the factorization theorem (\ref{eq:quantbetastart}).  Both $S^\mr{p}_q(t)$ and $\langle [\vec{r}_i(t) - \vec{r}_j(0)]^2 \rangle$ can be determined independently in the simulation.  By comparing the results we are able to test to what extent the displacements can be considered as Gaussian distributed random variables.

A further step consists in expressing $\langle [\vec{r}_i(t) - \vec{r}_j(0)]^2 \rangle$ by the Rouse modes $\vec{X}_p(t)$, defined by \cite{Verdier1966},
\begin{equation}
\vec{X}_p(t) := \frac{1}{N} \sum_{i=1}^{N}\vec{r}_i(t)\cos\left[\frac{p\pi(i-1/2)}{N}\right] \;,
\end{equation}
where $p$ ($=0,1,\ldots \hspace{-0.1pt}, N-1$) denotes the mode index. For our systems the Rouse modes were analysed in \cite{BennemannPaulBaschnagel1999_Rouse}.  This analysis showed that the Rouse-mode correlation functions, $\langle\vec{X}_p(t) \cdot \vec{X}_q(0)\rangle$, are (to a very good approximation) orthogonal at all times, i.e., $\langle\vec{X}_p(t) \cdot \vec{X}_q(0)\rangle \propto \delta_{pq}$.  When using this property the displacements may be expressed as
\begin{equation}
\label{eq:rouse_displ}
\begin{split}
&\left\langle \bigl[ \vec{r}_i(t) - \vec{r}_j(0)\bigr]^2 \right\rangle = g_3(t) + 4\sum_{p=1}^{N-1} \left\langle \vec{X}_p^2(0) \right\rangle \\
&\times \left( \cos\left[\frac{p\pi(i-1/2)}{N}\right] - \cos\left[\frac{p\pi(j-1/2)}{N}\right] \right)^2 \\
&+ \; 8\sum_{p=1}^{N-1}\left\langle \vec{X}_p^2(0) \right\rangle \bigl[1-\Phi_{pp}(t)\bigr]\\
&\times \cos\left[\frac{p\pi(i-1/2)}{N}\right] \cos\left[\frac{p\pi(j-1/2)}{N}\right] \;,
\end{split}
\end{equation}
where $g_3(t)$ denotes the mean-square displacement of the chains' centre of mass, and 
\begin{equation}
\label{eq:def_rouse_correl}
\Phi_{pp}(t) := \frac{\left\langle\vec{X}_p(t) \cdot \vec{X}_p(0)\right\rangle}{\left\langle\vec{X}_p(0) \cdot \vec{X}_p(0)\right\rangle} \quad (p=1,\ldots,N-1)
\end{equation}
are the normalized time-correlators of the Rouse modes.  Again, both sides of equation~(\ref{eq:rouse_displ}) can be obtained independently from the simulation.  We call the Gaussian approximation (\ref{eq:gauss_approx}) with displacements calculated via equation~(\ref{eq:rouse_displ}) ``Rouse ap\-pro\-xi\-ma\-ti\-on''.

\section{Simulation Results}
\label{sec:results}

\subsection{Statics}
\label{subsec:statics}
For the interpretation of the intermediate scattering functions it is important to know the static structure factors $S_q$ and $S^\mr{p}_q$.  $S_q$ was computed in \cite{BennemannBaschnagelPaul1999_incoherent} and is compared with $S^\mr{p}_q$ in Figure~\ref{fig:static_structure_factor_melt+chain+Debye}. Both $S_q$ and $S^\mr{p}_q$ exhibit an amorphous halo at $q_\mr{max} \approx 7$ corresponding to the nearest neighbour shell.  As $q$ grows, $S_q$ oscillates around 1, approaching 1 for the highest accessible $q$.  These features are characteristic of an amorphous structure and present at all temperatures.  With decreasing temperature the maxima and minima become sharper, and the position of the first maxima shifts to slightly larger $q$, as the melt becomes denser.

At small $q$ the structure factor of a chain, $S^\mr{p}_q$, can be fairly well described by the Debye function \cite{DoiEdwards}
\begin{eqnarray}
\label{eq:SpDebye}
S^\mr{p}_\mr{Debye}(q) &=& \frac{2N}{q^4 R_\mr{g}^4}\left( \mr{e}^{-q^2 R_\mr{g}^2} + q^2 R_\mr{g}^2 -1 \right) \\
\label{eq:SpDebye_Approx}
&\approx & \frac{N}{1 + q^2 R_\mr{g}^2/2} \;
\end{eqnarray}
Equation~(\ref{eq:SpDebye_Approx}) represents an approximation to the Debye function~(\ref{eq:SpDebye}), which is accurate to about 15\% \cite{DoiEdwards}.  The Debye function assumes a Gaussian distribution of the monomer-monomer distances, an assumption which is certainly not justified at higher $q$ probing smaller distances where the exact form of the potentials matters.  Therefore, deviations between simulation and theory are expected for $q > 2\pi/R_\mr{g}$.  At $q \approx 7.5 $  the simulated $S^\mr{p}_q$ has a maximum caused by the nearest neighbour shell.  Compared to $S_q$ it is shifted to larger $q$ because the distance between bonded monomers in a polymer is shorter than between non-bonded nearest neighbours. $S^\mr{p}_q$ varies only very slightly in the studied temperature range.  Thus, the data at $T=0.46$ are a representative example.

\bfig
\rsfig{figures/static_structure_factor_melt+chain+Debye.eps}
\caption[]{
\label{fig:static_structure_factor_melt+chain+Debye}
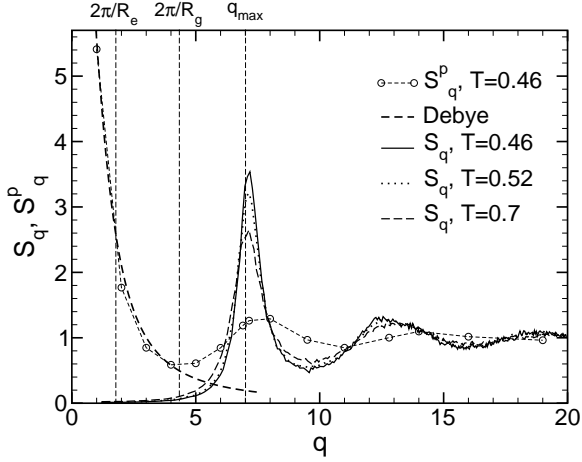
Static structure factors of the melt $S_q$ and of the chain $S^\mr{p}_q$.  $S_q$ is shown at three temperatures covering the range of investigated temperatures $\Tc \simeq 0.45 < 0.46 \le T \le 0.7$.  $S_q$ depends only weakly on temperature and exhibits typical features of a liquid structure at all $T$.  $S^\mr{p}_q$ remains unchanged in this temperature region and is shown together with the Debye approximation (\ref{eq:SpDebye}).  Reciprocal lengths corresponding to the end-to-end distance $R_\mr{e} \simeq 3.5$ and the radius of gyration $R_\mr{g} \simeq 1.45$ are also indicated.
}
\efig

\begin{figure}
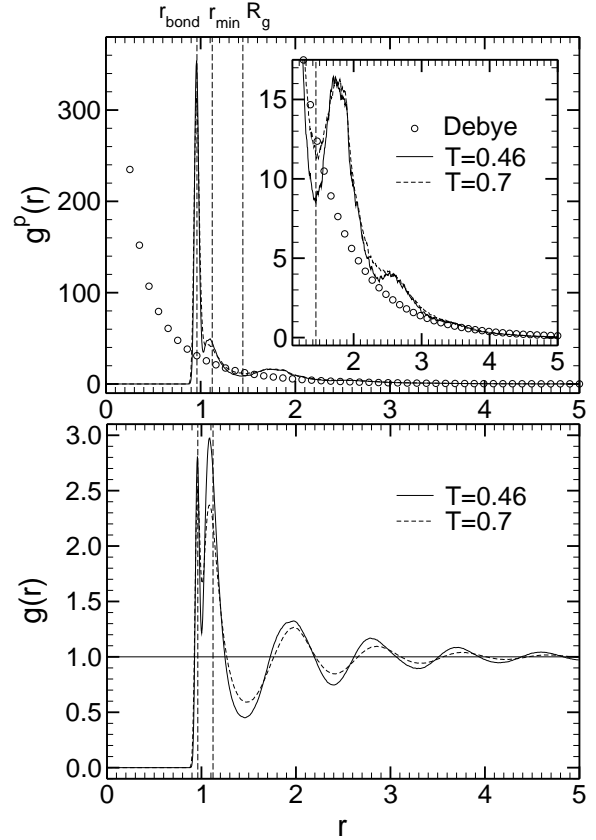

\begin{center}
\rsfig{figures/paircorrel_chain+Debye+melt_T0.46_T0.7.eps}
\caption[]{
\label{fig:paircorrel_chain+Debye+melt_T0.46_T0.7.eps}
Pair-correlation functions of the chain (top) and of the melt (bottom) for the lowest temperature $T=0.46$ and the highest temperature $T=0.7$.  The inset in the upper panel enlarges the region of $r \ge 1.1$.  The vertical lines are at $r_\mr{bond}=0.9609$, the preferred bond length, at $r_\mr{min} = 1.1225$, the preferred distance of non-bonded monomers, and at $r = R_\mr{g} = 1.45$, the radius of gyration.  The circles in the top panel represent the approximation (\ref{eq:gpDebye_Approx}) for $g^\mr{p}_\mr{Debye}(r)$.
}
\end{center}
\end{figure}

The corresponding quantities in real space, the pair correlation functions of the melt and the chain, $g(r)$ and $g^\mr{p}(r)$, are shown in Figure~\ref{fig:paircorrel_chain+Debye+melt_T0.46_T0.7.eps}.  The main features remain unchanged in the whole temperature range, but at lower temperature maxima and minima become more pronounced.  The overall appearance resembles closely the pair correlation function of hard spheres and simple Lennard-Jones liquids (see e.g.\ \cite{HansenMcDonald}).

However, there are also characteristic differences.  In a simple liquid there is only a single nearest neighbour peak, whereas the polymer model exhibits two peaks: one at $r_\mr{bond} = 0.9609$ and another one at $r_\mr{min} = 1.1225$ corresponding to the two length scales of the model \cite{BBPB_2000,BennemannBaschnagelPaul1999_incoherent}.  The peak at $r_\mr{min}$ reflects the sharp minimum of the bond potential.  The minimum of the Lennard-Jones potential is less steep.  Hence, one expects a wider peak.  Only monomers in the same chain which are not direct neighbours can contribute to the peak at $r_\mr{min}$ for $g^\mr{p}(r)$.  Therefore, it is much lower than that of $g(r)$ where surrounding polymers can also contribute.  The subsequent next-nearest shells cause the periodic sequence of maxima and minima in $g(r)$, whose amplitude gradually decreases as $r$ grows.  There is also a weak, broad second neighbour-shell maximum in $g^\mr{p}(r)$ around $r=1.8$. It is situated at smaller $r$ than the corresponding maximum of $g(r)$ at $r=1.95$, presumably because next neighbours are closer in the chain than in the melt, roughly by $r_\mr{min} - r_\mr{bond} = 0.1616$.  The inset in the plot for $g^\mr{p}$ shows a magnification of the comparison with the Fourier-transform of the approximation (\ref{eq:SpDebye_Approx}) for the Debye function ($V$ is the volume of the simulation box) 
\begin{equation}
\label{eq:gpDebye_Approx}
g^\mr{p}_\mr{Debye}(r) \approx \frac{2N}{(2\pi)^{3/2}(N-1)}
\frac{V}{R^2_\mr{g}}\,\frac{1}{r}\exp\Bigg(-\frac{\sqrt{2}r}{R_\mr{g}}\Bigg) \;.
\end{equation}
Equation~(\ref{eq:gpDebye_Approx}) describes $g^\mr{p}(r)$ quite well for $r \gtrsim 3$, where details of the monomer-monomer interactions do not matter.

\subsection{Dynamics}
\label{dyn}

\subsubsection{Intra-chain Coherent Scattering Function and Rouse Analysis}
\label{sp+rouse}

\bfig
\rsfig{figures/gaussian_approx_q_1_2_3_4_5_69_95_14_19_T065.eps}
\caption[]{
\label{fig:gaussian_approx_q_1_2_3_4_5_69_95_14_19_T065.eps}
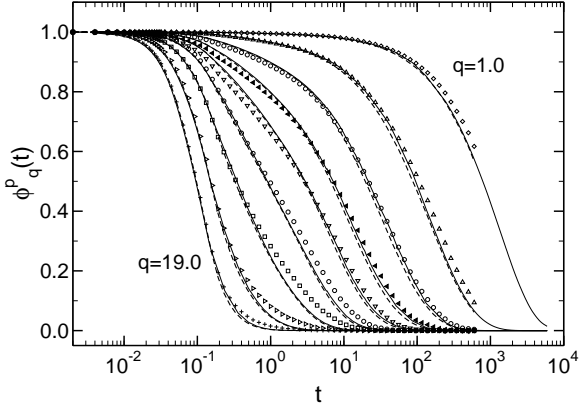
Coherent scattering function $\phi^\mr{p}_q(t)$ at $T=0.65$ for wave vector moduli $q=1$, 2, 3, 4, 5, 6.9, 9.5, 14, and 19.  Symbols show the simulation results. Dashed lines represent the Gaussian approximation (Eq.~(\ref{eq:gauss_approx})), solid lines the Rouse approximation, i.e., Eq.~(\ref{eq:gauss_approx}) together with Eqs.~(\ref{eq:rouse_displ}) and (\ref{eq:def_rouse_correl}). 
}
\efig

\bfig
\rsfig{figures/gaussian_approx_q_1_2_3_4_5_69_95_14_T048.eps}
\caption[]{
\label{fig:gaussian_approx_q_1_2_3_4_5_69_95_14_T048.eps}
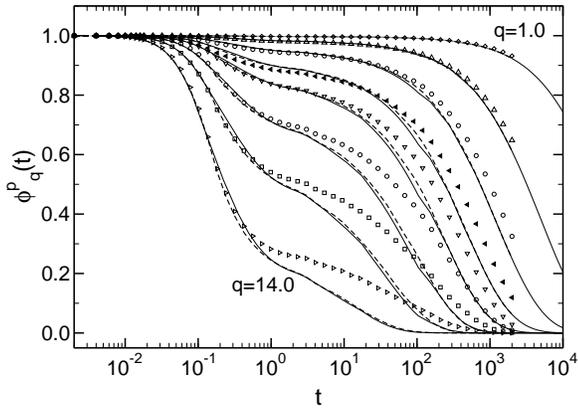
Same plot as in Figure~\ref{fig:gaussian_approx_q_1_2_3_4_5_69_95_14_19_T065.eps}, but at $T=0.48$ and for $q=1$, 2, 3, 4, 5, 6.9, 9.5, and 14.  The simulation results are represented by the symbols, whereas the dashed and solid lines show the Gaussian (Eq.~(\ref{eq:gauss_approx})) and Rouse approximations (Eq.~(\ref{eq:gauss_approx}) with Eqs.~(\ref{eq:rouse_displ}) and (\ref{eq:def_rouse_correl})), respectively.
}
\efig

In this section we compare the Gaussian and Rouse approximations (see Section~\ref{subsubsec:rouse}) with the simulation data for the intra-chain coherent scattering function at different $q$.  For $T=0.65$ Figure~\ref{fig:gaussian_approx_q_1_2_3_4_5_69_95_14_19_T065.eps} shows that both approximations agree with one another and describe the correlators quite well.  Deviations occur for $q=4$ and 5 around $t=0.5$ and for all $q$, especially for $q \geq 6.9$, in the $\alpha$-regime.  These observations suggest the following conclusions:  The agreement between the Gaussian and the Rouse approximations shows that equation~(\ref{eq:rouse_displ}) represents an accurate description of $\langle [\vec{r}_i(t) - \vec{r}_j(0)]^2 \rangle$.  This means that the Rouse modes are, to a very good approximation, orthogonal at all times for our model at high temperatures.  Thus, the difference between simulation and theory must be attributed to non-Gaussian distributed displacements $\vec{r}_i(t) - \vec{r}_j(0)$.  The influence of this non-Gaussian character depends on $q$ and $t$, but in general makes the simulated scattering functions decay more slowly than the theory.  The same conclusion was also drawn from atomistic simulations of polyethylene \cite{paul1998} and polybutadiene \cite{GrantPaul2000,GrantPaul_ChemPhys2000}.

At $T=0.48$, both the Gaussian and Rouse approximations poorly describe the $\alpha$-relaxation for $q \geq 3$, and fail completely for large $q$ already at the beginning of the $\beta$-regime (see Figure~\ref{fig:gaussian_approx_q_1_2_3_4_5_69_95_14_T048.eps}).  Although a two-step relaxation behaviour is reproduced, the collective effects in the polymer are not properly taken into account.  Since the Gaussian approximation fails, it can be concluded that the forces acting on the monomers do not give rise to Gaussian distributed displacements as temperature approaches $\Tc$. On the other hand, the figures show that both the Gaussian and Rouse approximations are almost indistinguishable at all $q$.  Thus, the Rouse-mode formula (\ref{eq:rouse_displ}) for the displacements remains accurate even in the supercooled regime close to $\Tc$ (this was also checked by comparing the displacements from simulation data directly with equation~(\ref{eq:rouse_displ})).

\subsubsection{Dynamics of the melt in the $\beta$-relaxation regime in $q$-space}
\label{subsubsec:beta_q}
We want to test the validity of the factorization theorem (cf.\ equation~(\ref{eq:quantbetastart}))
\begin{equation}
\label{eq:factorization_q}
\phi^\mr{x}_q(t) = f^\mr{x c}_q + \frac{h^\mr{x}_q t_0^{\,a}}{t_\sigma^{\,a}}g(t / t_{\sigma})
\end{equation}
by computing the function \cite{GleimKob2000}
\begin{equation}
\label{eq:def_R}
R^\mr{x}_q(t) := \frac{\phi^\mr{x}_q(t) - \phi^\mr{x}_q(t')}{\phi^\mr{x}_q(t'') - \phi^\mr{x}_q(t')} = \frac{g(t/ t_{\sigma}) - g(t'/ t_{\sigma})}{g(t''/ t_{\sigma}) - g(t'/ t_{\sigma})} \equiv R(t) \;.
\end{equation}
If the factorization property holds, $R^\mr{x}_q(t)$ depends only on the fixed times $t'$ and $t''$ and on temperature (via $t_\sigma$), but not on the correlator ``x'' or on $q$.  The times $t'$ and $t''$ can be chosen arbitrarily in the $\beta$-regime.   For numerical stability it is, however, advisable to take values at the beginning and the end of the plateau region in order to obtain a large denominator.  An advantage of equation~(\ref{eq:def_R}) is that it represents a simple test which works directly with the simulation data.  No intricate fit procedure is involved.  This approach has therefore been pursued in reciprocal and in real space in several other studies \cite{GleimKob2000,SignoriniBarratKlein1990,KobAndersen_LJ_I_1995} (see also \cite{ToelleSchoberWuttke1997,WuttkeSeidl1998} for comparable experimental tests).

Figure~\ref{fig:R_A_T046} shows $R^\mr{x}_q(t)$ computed from the scattering functions at $T=0.46$ with $t''=0.610$ and $t'=86.43$.  This temperature is already so close to $\Tc$ ($\simeq 0.45$) that deviations from the ideal MCT prediction in the $\alpha$-regime, i.e., from the power law behaviour (\ref{eq:alpha_time_law}) of the $\alpha$-timescale, occur \cite{BennemannBaschnagelPaul1999_incoherent,BennemannPaulBaschnagel1999,BennemannPaulBaschnagel1999_Rouse,alphaDynamics}.  Such deviations are also found in other simulations \cite{KaemmererKobSchilling1998,KaemmererKobSchilling1998_orient,KaemmererKobSchilling1997} or in experiments, see e.g.\ \cite{LunkenheimerReview2000}, and are usually attributed to additional relaxation channels which are not taken into account by idealized MCT, but become dominant as temperature decreases towards and below $\Tc$ \cite{GoetzeSjoegren1995_TTSP,FuchsGoetzeHildebrand1992_extMCT}.  Nonetheless, Figure~\ref{fig:R_A_T046} illustrates that there is an intermediate time-window of about 2.5 decades where the data for incoherent and coherent scattering collapse onto a $q$-independent master curve, while they splay out at both short and late times.  We tested that the master curve is the same for all correlators ``x''.  Very similar results are obtained for higher temperatures, with times appropriately chosen in the plateau-region for each temperature, as long as a two-step relaxation is observed (i.e., for $T \lesssim 0.52$ \cite{BennemannBaschnagelPaul1999_incoherent}).  The observation that the factorization theorem is satisfied for temperatures where ergodicity restoring processes already violate the validity of equation~(\ref{eq:alpha_time_law}) suggests that it remains valid also for $T \lesssim \Tc$ (as expected theoretically \cite{FuchsGoetzeHildebrand1992_extMCT}).

Furthermore, Figure~\ref{fig:R_A_T046} qualitatively confirms predictions for the higher order corrections to the factorization theorem:  According to equation~(\ref{eq:B_A+A_A}), the short- and long-time correction coefficients have the same $q$-dependence.  These corrections become important before and after the central $\beta$-regime.  This means that the top curve before the collapse onto the master curve is also the top curve after the collapse, the second from top before $t''$ is the second from top after $t'$, and so on.  This behaviour is illustrated with dashed and dot-dashed lines in Figure~\ref{fig:R_A_T046} for some $q$.  Except for the coherent scattering functions, which exhibits oscillations at small $q$ in the early $\beta$-regime (see part II of this paper for a more detailed discussion of this point \cite{alphaDynamics}) the prediction is seen to be fulfilled.  One finds that the correction coefficients for the incoherent scattering function, $A^\mr{s}_q$ and $B^\mr{s}_q$, grow monotonously with $q$, in agreement with calculations for an ideal hard sphere system \cite{FuchsGoetzeMayr1998}.  Similar findings were reported in \cite{GleimKob2000} for the incoherent scattering function of a binary Lennard-Jones fluid. 

\bfig
\rsfig{figures/R_A_T046.eps}
\caption[]{
\label{fig:R_A_T046}
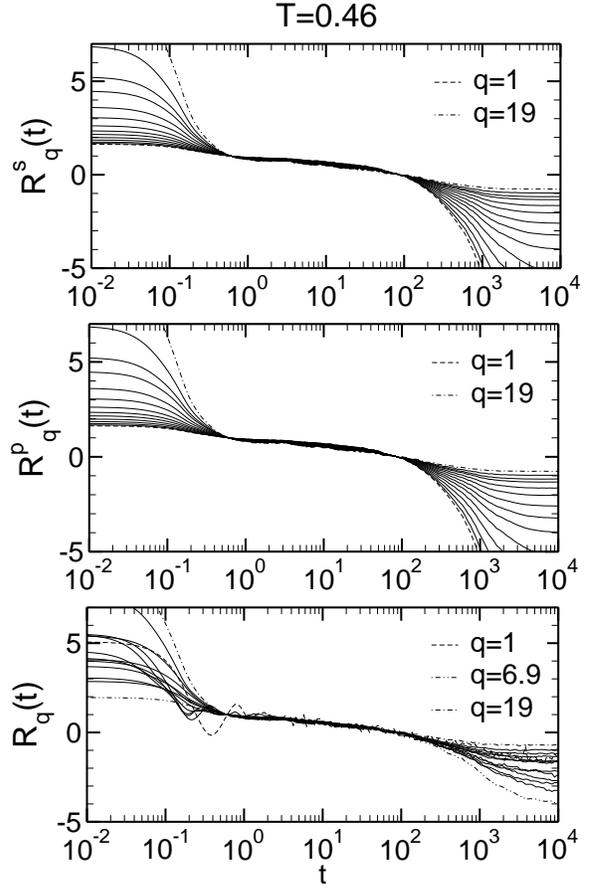
Test of the factorization theorem for all scattering functions at $T=0.46$ by plotting $R^\mr{x}_q(t)$, defined by equation~(\ref{eq:def_R}), vs.\ $t$.  From top to bottom the three panels show the results for incoherent, coherent-chain and coherent-melt scattering.  In the $\beta$-regime all curves collapse onto a single master curve.  By definition, $R^\mr{x}_q(t'' = 0.610) = 1$ and $R^\mr{x}_q(t' = 86.43) = 0$.
}
\efig

\subsubsection{Dynamics of the Rouse modes in the $\beta$-relaxation regime}
\label{subsubsec:beta_Rouse}
Motivated by Figure~\ref{fig:R_A_T046} it is tempting to apply equation~(\ref{eq:def_R}) also to polymer specific quantities, such as the Rouse mo\-des.  Therefore, we define $R^\mr{Rouse}_p(t)$ in analogy to $R^\mr{x}_q(t)$ with $\phi^\mr{x}_q(t)$ replaced by $\Phi_{pp}(t)$ (see Eq.~(\ref{eq:def_rouse_correl})).  Figure~\ref{fig:R_A_from_Rouse-modes_interpol+inc_sf_T048} shows $R^\mr{Rouse}_p(t)$ (for all Rouse modes $p$) and compares it to $R^\mr{s}_q(t)$ at $T=0.48$.  Note the different scale for the y-axis, causing the spread of $R^\mr{s}_q(t)$ to appear larger than in Figure~\ref{fig:R_A_T046}.   Again, $t''$ and $t'$ are chosen at the beginning and the end of the plateau.  

As with the scattering functions, all Rouse correlators collapse onto a master curve in the $\beta$-plateau region.  Although the splaying out at short and long times is much weaker than for the incoherent scattering function, the figure suggests that the factorization property also holds for the Rouse modes.  However, the master curve of $R^\mr{Rouse}_p(t)$ lies slightly above that of $R^\mr{x}_q(t)$.  Hence, one can speculate that an extension of MCT to polymer melts might yield a $\beta$ master function for the Rouse modes, which is close, but perhaps not identical to $g(t)$. 

On the other hand, this conjecture has to be considered with some reservations, since the Rouse modes exhibit a two-step relaxation with a rather high plateau value (the non-ergodicity parameter is presumably larger than 0.9 for all $p$) \cite{BennemannPaulBaschnagel1999_Rouse}.  This proviso is suggested by a recent comparative analysis of depolarized-light scattering, dielectric-loss and incoherent neutron scattering results for propylene carbonate \cite{goetzevoigtmann2000}.  The study shows that the data from these different techniques can be consistently described by a schematic MCT model which goes beyond asymptotic predictions and incorporates corrections resulting from hopping processes and from the crossover to vibrations and to the $\alpha$-process.  By comparing this description with an analysis using only the asymptotic laws it was found that the leading-order results can be strongly masked for quantities with large non-ergodicity parameters.  

Finally, Figure~\ref{fig:R_A_from_Rouse-modes_interpol+inc_sf_T048} allows another observation.  It reveals a (further) difference between $R^\mr{Rouse}_p(t)$ and $R^\mr{s}_q(t)$.  The corrections to the factorization property for the Rouse modes behave oppositely to those of $R^\mr{s}_q(t)$: The order from top to bottom of the $R^\mr{Rouse}_p(t)$ curves before $t''$ is reversed after $t'$ for all $R^\mr{Rouse}_p(t)$ (only shown for $p=1$), whereas it is maintained for the scattering functions (see also Figure~\ref{fig:R_A_T046}). 

\bfig
\rsfig{figures/R_A_from_Rouse-modes_interpol+inc_sf_T048.eps}
\caption[]{
\label{fig:R_A_from_Rouse-modes_interpol+inc_sf_T048}
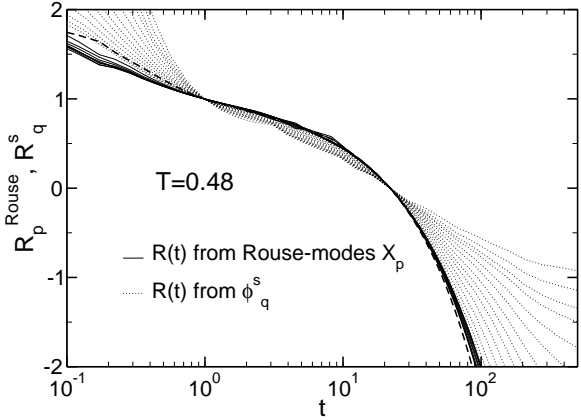
$R^\mr{s}_q(t)$ and $R^\mr{Rouse}_p(t)$ for $T=0.48$.  $R^\mr{Rouse}_p(t)$ is defined analogously to equation~(\ref{eq:def_R}) by replacing $\phi_q^\mr{x}(t)$ by the Rouse modes $\Phi_{pp}(t)$.  The times $t'$ and $t''$ are chosen as $t'' = 0.988$ and $t' =  21.97$.  The order from top to bottom of the $R^\mr{s}_q(t)$ curves outside the collapse region is conserved (see Figure~\ref{fig:R_A_T046}), whereas the order of the $R^\mr{Rouse}_p(t)$ curves is reversed (exemplified for $p=1$ drawn as a thick dashed line).
}
\efig

\subsubsection{Dynamics of the melt in the $\beta$-relaxation regime in real space}
\label{subsubsec:beta_r}

\bfig
\rsfig{figures/Hr_over_Hrp_all_vanhove_T048.eps}
\caption[]{
\label{fig:Hr_over_Hrp_all_vanhove_T048}
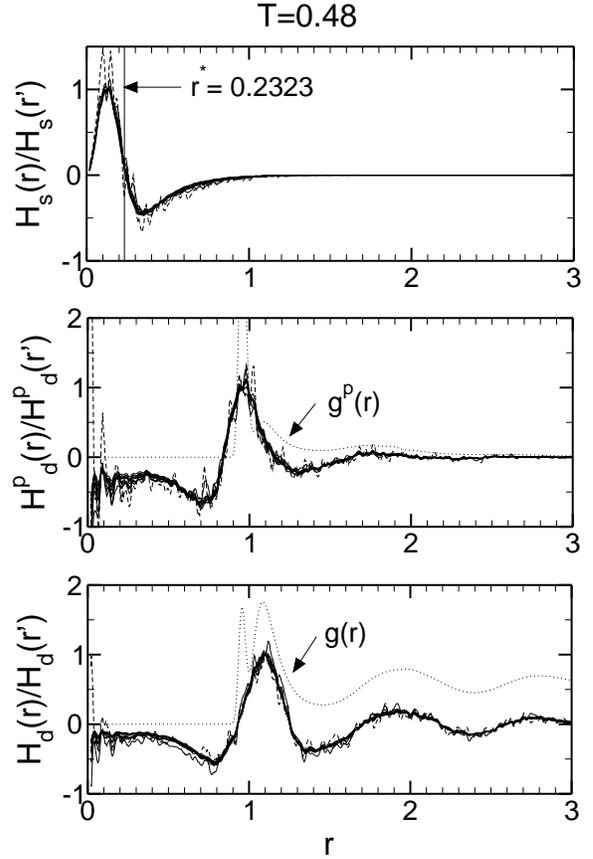
Test of the factorization theorem for the van Hove correlators at $T=0.48$.  $\tilde{R}_\mr{X}(r,t)$ is depicted for 9 times from the interval $t''=0.988 \leq t < t'=21.97$ (same values for $t''$ and $t'$ as in Figure~\ref{fig:R_A_from_Rouse-modes_interpol+inc_sf_T048}).   For the distinct correlators the respective pair correlation functions (rescaled) are shown as dotted lines for comparison.  $r^*$ is the zero of $\tilde{R}_\mr{s}(r,t)$.  For $0.46 \le T \le 0.52$ all $\tilde{R}_\mr{X}(r,t)$ fall on a temperature independent master curve for each correlator.  This master curve is the same for all temperatures.  In order not to overload the figure only the results for $T=0.48$ are shown.  The dashed lines correspond to $t$ closest to $t'$ where one can see numerical instabilities (note that $\tilde{R}_\mr{X}(r,t)$ is undetermined for $t=t'$). 
}
\efig

The factorization property can as well be investigated in real space for the van Hove functions, for which one obtains by Fourier transformation of equation~(\ref{eq:factorization_q})
\begin{equation}
\label{eq:factorization_r}
G_\mr{X}(r,t) = F_\mr{X}(r) + (t_0/t_\sigma)^a H_\mr{X}(r) g(\hat{t}) \;.
\end{equation}
We now define 
\begin{equation}
\label{eq:def_RX}
\tilde{R}_\mr{X}(r,t) := \frac{G_\mr{X}(r,t) - G_\mr{X}(r,t')}{G_\mr{X}(r',t) - G_\mr{X}(r',t')} = \frac{H_\mr{X}(r)}{H_\mr{X}(r')} \equiv \tilde{R}_\mr{X}(r) \;,
\end{equation}
which was proposed in \cite{SignoriniBarratKlein1990,KobAndersen_LJ_I_1995}.  We can choose any time $t'$ in the $\beta$-regime and an arbitrary value for $r'$.  A possibility to fix $r'$ is to take the position of the nearest neighbour maximum at $t=0$ for the distinct correlators and that of the maximum of $4\pi r^2 G_\mr{s}(r,t)$ in the centre of the $\beta$-regime for the self correlator: $r^{\prime\, \mr{p}}_\mr{d} = 0.9575$, $r^{\prime}_\mr{d} = 1.1025$, and $r'_\mr{s} = 0.13253$.  Since the correlators decay faster at the peak positions than at other $r$, the denominator of equation~(\ref{eq:def_RX}) is largest at $r'_\mr{X}$.  This is again favourable for numerical stability. Furthermore, $\tilde{R}_\mr{X}(r'_\mr{X},t) = 1$ by definition.  If the factorization property is satisfied, the results for $\tilde{R}_\mr{X}(r, t)$ should fall onto a master curve at any time $t$ in the window of the $\beta$-relaxation.  The master curve is specific for each correlator, but independent of $T$, as the critical amplitude $H_\mr{X}(r)$ does not depend on temperature.  

We calculated $\tilde{R}_\mr{X}(r,t)$ from the van Hove functions at various $T$.  As an example, Figure~\ref{fig:Hr_over_Hrp_all_vanhove_T048} shows the results for $T=0.48$, which is the centre of the temperature range where a quantitative $\beta$-analysis was performed (see section~\ref{subsubsec:quantbeta} and \cite{BennemannBaschnagelPaul1999_incoherent}).  As expected from the factorization theorem there is a  characterisitic, temperature independent master curve for each correlator.  The distinct correlators $H_\mr{d}(r)$ and $H^\mr{p}_\mr{d}(r)$ roughly follow the respective pair correlation functions if $r\gtrsim 1$ (= monomer diameter).  Comparable results are also obtained in reciprocal space (see section~\ref{subsubsec:quantbeta}).  The spatial dependence of and the differences between different correlators can however be seen more clearly in real than in $q$-space (compare Figures~\ref{fig:Hr_over_Hrp_all_vanhove_T048} and \ref{fig:crit_ampl_vs_q}).  The behaviour of $\tilde{R}_\mr{d}(r,t)$ for our system is qualitatively similar to that of a binary Lennard-Jones mixture \cite{KobAndersen_LJ_I_1995,NaurothKob1997} and of hard spheres \cite{BarratGoetzeLatz1989}.  It is important to note that the master curves of Figure~\ref{fig:Hr_over_Hrp_all_vanhove_T048} are only obtained for times $t'' \leq t < t'$ chosen from the intermediate time-window of the plateau (see Figures~\ref{fig:R_A_T046} and \ref{fig:R_A_from_Rouse-modes_interpol+inc_sf_T048}).  If one leaves this time-window towards microscopic times or towards late times of the $\alpha$-process, the superposition of the data is no longer possible.  The same observation was also made for the binary Lennard-Jones mixture \cite{KobAndersen_LJ_I_1995}.

Due to the factorization property the spatial variation of $H_\mr{X}(r) / H_\mr{X}(r')$ provides information about the length scales which are involved in the $\beta$-relaxation.  Figure~\ref{fig:Hr_over_Hrp_all_vanhove_T048} illustrates that all $H_\mr{X}(r)$ quickly vanish on the scale of a few interparticle distances.  The most long-ranged decay is found for the critical amplitude of the melt's distinct correlator $H_\mr{d}(r)$ whose oscillations are damped out if $r \gtrsim 4$.  Thus, the $\beta$-relaxation involves monomer rearrangements up to about the forth nearest neighbour shell.  The dominant contribution comes from cooperative displacements of a monomer and its first and second neighbours.  This local character of the spatial variation of $H_\mr{X}(r) / H_\mr{X}(r')$ is an illustration of the ``cage effect'' \cite{Goetze1999_review,goetzemctessentials,GoetzeSjoegren1995_TTSP,Goetze_LesHouches}. 

Let us consider the spatial variation of $H_\mr{X}(r) / H_\mr{X}(r')$ in some more detail.  The critical amplitude of the self correlator $H_\mr{s}(r)$ has a root at $r^* = 0.2323$. This value is (almost) identical to $\sqrt{6} r_\mr{sc} = 0.2327$, i.e., to the root of the approximation,
\begin{equation}
\label{eq:big_r_H}
\frac{H_\mr{s}(r)}{H_\mr{s}(r^\prime)} \approx
\frac{(1-r^2/6r^2_\mr{sc}) \exp\big(-r^2/4 r^2_\mr{sc}\big)}
{(1-r^{\prime\, 2}/6r^2_\mr{sc}) \exp\big(-r^{\prime\, 2}/4 r^2_\mr{sc}\big)}\;,
\end{equation}
where $r_\mr{sc} = 0.095 \pm 0.005$ \cite{BennemannBaschnagelPaul1999_incoherent}.  Equation~(\ref{eq:big_r_H}) is obtained by Fourier transformation of a Gaussian approximation for the critical amplitude $h_q^\mr{s}$ (see Eq.~(31b) of \cite{FuchsGoetzeMayr1998}).  This approximation provides a reasonable description of the simulation data for $h_q^\mr{s}$ if $q < q_\mr{max}$ \cite{BennemannBaschnagelPaul1999_incoherent}.  Although it cannot be quantitatively precise when $r\lesssim  2\pi/q_\mr{max} \approx 0.9$, it qualitatively reproduces the features of $H_\mr{s}(r)$ for $r\geq r^*$ (root at $\sqrt{6}r_\mr{sc}$, minimum around $\sqrt{10}r_\mr{sc} \simeq 0.3$) and proposes that the relevant length scale for the monomer motion in the $\beta$-regime is $r_\mr{sc}$.  Numerically, the magnitude of $r_\mr{sc}$ ($=0.095$) is comparable to the Lindemann melting criterion, stating that a (crystalline) solid melts if a particle moves on average more than 0.1 of its own diameter around its equilibrium position.  This suggests that the local motion of a monomer in the $\beta$-regime resembles that of a particle in an amorphous solid which is about to melt.  

A similar conclusion was also drawn from a detailed analysis of the mean-first passage time in simulations of a binary LJ-mixture of small and large particles \cite{Allegrini1999}.  If temperature is close to $\Tc$, the mean-first passage time of a large particle to reach a distance $r$ from its origin grows most steeply for $0.21 \lesssim r \lesssim 0.25$.  Furthermore, the distribution of first passage times exhibits a power-law decay for this range of $r$ (see Figs.~8 and 9 of \cite{Allegrini1999}).  These features are indicative of intermittency in particle motion for distances which are close to $r^*$ of the binary mixture (see Fig.~7b of \cite{KobAndersen_LJ_I_1995}) and of our model.  Therefore, the analysis of Ref.~\cite{Allegrini1999} supports the interpretation that the vanishing of $H_\mr{s}(r)$ at $r^*$ defines a length scale $r_\mr{sc}$ for transient particle localization in an amorphous structure, as also suggested in \cite{Goetze_LesHouches}, for instance.  

Figure~\ref{fig:Hr_over_Hrp_all_vanhove_T048} shows that $H_\mr{s}(r)$ is positive if $r < r^*$, but negative for $r > r^*$.  Due to equation~(\ref{eq:factorization_r}) the product $H_\mr{s}(r) g(\hat{t})$ determines $G_\mr{s}(r,t)$, i.e., the probability to find monomer displacements of size $r$ in time $t$.  Since the $\beta$ master function $g(\hat{t})$ decreases monoto\-nous\-ly \cite{FranoschFuchsGoetze1997}, we can write for the rate by which the van Hove functions decay with time 
\begin{equation}
\label{eq:GxDecay}
\frac{\partial G_\mr{X}(r, t)}{\partial t} = - (t_0/t_\sigma)^a H_\mr{X}(r) \left |\frac{\partial g(\hat{t})}{\partial t} \right | \propto  - H_\mr{X}(r) \;.
\end{equation}
Therefore, the probability for monomer displacements of size $r$ in the $\beta$-regime decreases most at the maximum position of $H_\mr{s}(r)$ ($r \approx 0.13$) and increases most at its minimum position ($r \approx 0.35$), whereas it vanishes if $r \rightarrow 1$ (= monomer diameter).  On the level of the individual monomer motion this should imply that a monomer is partially reflected back to its origin when attempting to penetrate into the zone initially excluded by its neighbours.  This ``caging'' has been nicely demonstrated in a simulation of a (polydisperse) hard sphere system \cite{DoliwaHeuer1998} by calculating the probability that a particle moves parallel to the direction of its preceeding displacement.  It was found that this motion is on average oriented opposite to the initial direction if $r \lesssim 0.8$.   

The dynamics of the distinct correlators $H_\mr{d}^\mr{p}(r)$ and $H_\mr{d}(r)$ can also be interpreted by equation~(\ref{eq:GxDecay}).  Since $H_\mr{d}^\mr{p}(r)$ and $H_\mr{d}(r)$ are in phase with the respective pair-correlation functions if $r \gtrsim 1$ (Figure~\ref{fig:Hr_over_Hrp_all_vanhove_T048}),  equation~(\ref{eq:GxDecay}) illustrates that the van Hove correlators $G_\mr{d}^\mr{p}(r,t)$ and $G_\mr{d}(r,t)$ change most at the positions where the probability to find another particle at $t=0$ was largest or lowest.  The initial positions of high distinct-monomer density are strongly depleted, whereas the initial exclusion zones are populated.  So, other monomers also partially  enter the region $r < 1$ that a monomer occupied at $t=0$.  In this region the behavior of $H_\mr{d}^\mr{p}(r)$ and $H_\mr{d}(r)$ resembles a mirror image of the shape of $H_\mr{s}(r)$ for $r \gtrsim r^*$.  This is not unreasonable.  If the probability for a monomer displacement increases particularly at a distance $r \approx 0.35$ from its initial position, the approach of other particles towards this position should be most pronounced around $r \approx 0.7$ due to the preferred intermonomer distances $r_\mr{bond}=0.9609$ and $r_\mr{min}=1.1225$, assuming a nearest neighbour replaces this particle.  The comparison of $H_\mr{d}^\mr{p}(r)$ and $H_\mr{d}(r)$ further suggests that bonded nearest neighbours penetrate the initial exclusion zone of a monomer more easily than non-bonded ones as $H^\mr{p}_\mr{d}(r) < H_\mr{d}(r)$ for $r < 1$.  This should imply that adjacent monomers along the backbone of a chain tend to follow each other.  On the other hand, the small values of both $|H_\mr{d}^\mr{p}(r)|$ and $|H_\mr{d}(r)|$ for $r \lesssim 0.7$ indicate that the replacement of a monomer by its neighbours occurs only slowly with time in the $\beta$-regime.  The replacement process is slow because the monomer motion is in turn confined to displacements $r < 1$ by its neighbours.  This is another evidence for transient ``caging''.  A microscopic analysis of this monomer motion is underway \cite{Aichele_DH2000}.

\subsubsection{Quantitative $\beta$-analysis}
\label{subsubsec:quantbeta}
In this section we turn to a detailed quantitative analysis of the $\beta$-process.  From previous work on the incoherent scattering function \cite{BennemannBaschnagelPaul1999_incoherent} a lot of information is available which we want to exploit as much as possible to extend the description to coherent scattering.  Our aim is to reduce any further fitting to a minimum.  The starting point of the analysis is equation~(\ref{eq:quantbetaformula}).  In \cite{BennemannBaschnagelPaul1999_incoherent} and in this work it was found that the asymptotic formulas for the description of the $\beta$-relaxation are applicable for $0.47 \le T \le 0.52$. Thus, this is the temperature range where the $\beta$-analysis is carried out.

The exponents $a$ and $b$, and the constant $B$ were determined in Ref.~\cite{BennemannBaschnagelPaul1999_incoherent},
\begin{equation}
a = 0.352 \pm 0.010, \; b = 0.75 \pm 0.04, \; B=0.476 \pm 0.060\;.
\end{equation}
Additionally, the $\beta$ master function $g(\hat{t})$ was numerically given, along with the $\beta$-times $t_\sigma$ for all temperatures.  Thus, the complete time and temperature dependences in the $\beta$-regime were known.  

Using the $\beta$ master function the condition $g(\hat{t}_\mr{co}) = 0$ allows us to determine the crossover time $t_\mr{co} = \hat{t}_\mr{co} t_\sigma$.  Now, we can read off the non-ergodicity parameters from the simulation data by
\begin{equation}
\phi^\mr{x}_q(t_\mr{co}) = f^\mr{x c}_q \;,
\end{equation}
since higher order corrections are vanishingly small at $t_\mr{co}$ \cite{FranoschFuchsGoetze1997,FuchsGoetzeMayr1998}.  

Theoretically, $f^\mr{x c}_q$ should not be temperature dependent, but we observed a slight dependence on $T$.  So, we averaged over all temperatures to obtain the values shown in Figure~\ref{fig:NEPs+S_q_vs_q}.  $f^\mr{sc}_q$ and $f^\mr{pc}_q$ are very close to one another, being different only at small $q$.  For $q \ge 11$ all curves fall on top of each other.  $f^\mr{c}_q$ is in phase with $S_q$, as theoretically predicted for hard spheres \cite{FranoschFuchsGoetze1997} and also found in simulations of a binary Lennard-Jones liquid \cite{KobAndersen_LJ_II_1995,GleimKobBinder1998} and diatomic molecules \cite{KaemmererKobSchilling1998}.  Interpreted within the framework of MCT this means that modes at the next neighbour distance (where $S_q$ has its maximum $q_\mr{max}$) decay very slowly and contribute strongly to the slowing down of the dynamics of the melt.  Interestingly, there is a weak shoulder at $q\approx 4.5$ corresponding approximately to the size of a chain, i.e., $2\pi /R_\mr{g} \approx 4.35$.  A similar shoulder was also found in simulations of diatomic molecules \cite{KaemmererKobSchilling1998}, but is absent for ideal hard spheres \cite{FranoschFuchsGoetze1997}.  An explanation could be that  the polymer coil may be viewed as a soft, diffuse particle with radius $R_\mr{g}$, which contributes on its length scale to the slowing down of structural relaxation.  The same conclusion can be drawn from the $\alpha$-relaxation times as discussed in part II \cite{alphaDynamics}.  This conjecture could be further tested by simulations with polymers of different length.  

\bfig
\rsfig{figures/NEPs+S_q_vs_q.eps}
\caption[]{
\label{fig:NEPs+S_q_vs_q}
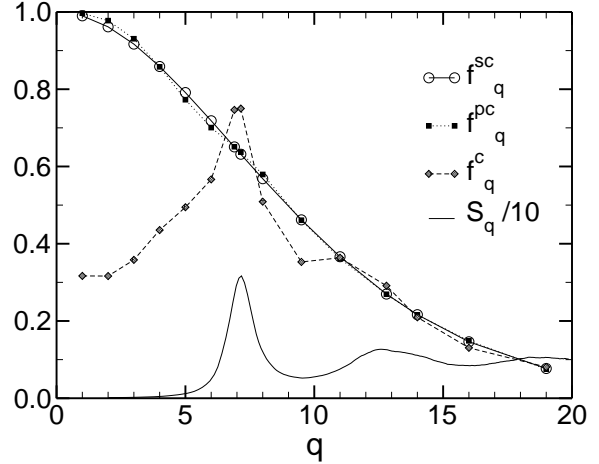
Non-ergodicity parameters $f^\mr{sc}_q$, $f^\mr{pc}_q$, and $f^\mr{c}_q$ versus $q$. $f^\mr{c}_q$ is in phase with the static structure factor, $S_q$, shown as a thin solid line.
}
\efig

In the vicinity of $t_\mr{co}$ higher order corrections are negligible and the critical amplitude (combined with a prefactor to $h^\mr{x \; fit}_q$, Eq.~(\ref{eq:def_h_fit})) can be obtained by numerically evaluating
\begin{equation}
h^\mr{x \; fit}_q = \frac{\partial_t \phi^\mr{x}_q(t)}{\partial_t[g(\hat{t}) / t^a_\sigma]} \biggr|_{t=t_\mr{co}} \;.
\end{equation}

Again, the curves for $h^\mr{s \; fit}_q$ and $h^\mr{p \; fit}_q$ are very similar, especially at large $q$ as shown in Figure~\ref{fig:crit_ampl_vs_q}.  The most prominent structure is observed for $h^\mr{fit}_q$.  In the ideal hard sphere system there is a sharp minimum at $q_\mr{max}$, whereas $h_q$ exhibts only slight variations for larger $q$, which are roughly in phase with $S_q$ \cite{FranoschFuchsGoetze1997}.  For diatomic molecules \cite{KaemmererKobSchilling1998} and water \cite{SciortinoFabbianChen1997} a distinct minimum was also found at $q_\mr{max}$. For $q < q_\mr{max}$, $h_q$ first passes through a (more or less pronounced) maximum in these systems before it decreases (and eventually levels off) if $q \rightarrow 0$.  This is in contrast to our system where $h_q$ first shows a minimum and then grows for $q \rightarrow 0$.  However, one has to be cautious with the results of coherent melt scattering at small $q$.  On the one hand, the statistics is worse than at larger $q$, and on the other hand, we observed oscillations at the beginning of the $\beta$-process in $\phi_q(t)$ in this $q$-range, which might be finite size effects (see Fig.~\ref{fig:Hr_over_Hrp_all_vanhove_T048} and part II \cite{alphaDynamics} for a more complete discussion).  Despite this proviso it is interesting to note that $h_q$ exhibits a minimum at $q=4$, which is close to the position where $f_q^\mr{c}$ has a shoulder.  So, one may speculate that these features reflect the slowing down of a chain much in the same way as the shoulder and the minimum at $q_\mr{max}$ signal that of particles in a non-polymeric liquid.  The minimum at $q_\mr{max}$ is visible here in form of a step at $q \approx 7$ only.  In contrast to the previously mentioned systems \cite{KaemmererKobSchilling1998,SciortinoFabbianChen1997,FranoschFuchsGoetze1997} we find a clear maximum at $q \approx 8 > q_\mr{max}$. For $q \ge 8$, the curve oscillates with $S_q$ (around $h^\mr{s \; fit}_q$), which is qualitatively in accord with results for hard spheres \cite{FranoschFuchsGoetze1997}.  

\bfig
\rsfig{figures/crit_ampl_vs_q.eps}
\caption[]{
\label{fig:crit_ampl_vs_q}
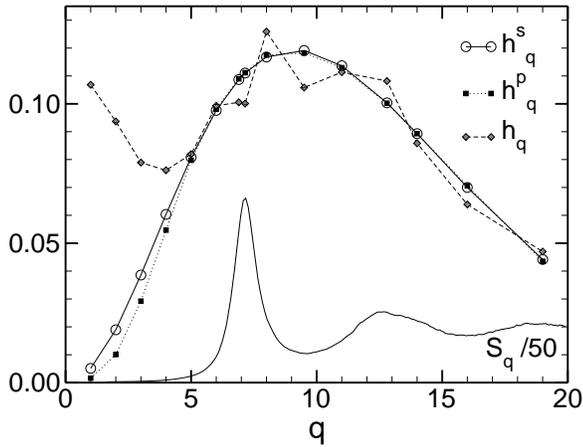
Critical amplitudes $h^\mr{s \; fit}_q$, $h^\mr{p \; fit}_q$, and $h^\mr{fit}_q$ (Eq.~(\ref{eq:def_h_fit})) together with the static structure factor $S_q$ (thin solid line) versus $q$. 
}
\efig

\bfig
\rsfig{figures/long_time_coeff_vs_q.eps}
\caption[]{
\label{fig:long_time_coeff_vs_q}
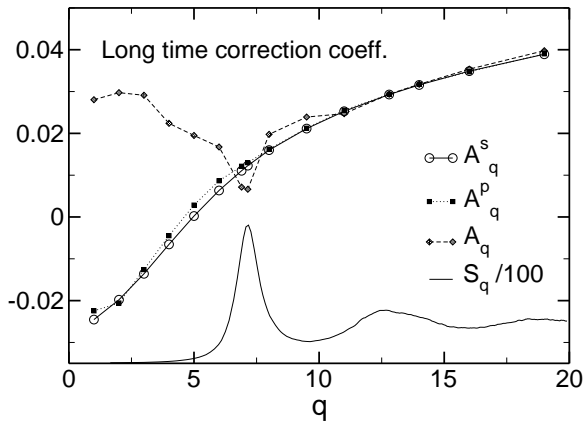
Long-time correction coefficients $A^\mr{s}_q$, $A^\mr{p}_q$, and $A_g$ (Eqs.~(\ref{eq:def_A_A}) and (\ref{eq:def_B_A})). The static structure factor $S_q$ is shown as a solid line for comparison.  
}
\efig

Up to now, we have not considered higher order corrections to the factorization theorem, that is, the third and forth terms of equation~(\ref{eq:quantbetaformula}). They were found to be quite significant outside the centre of the $\beta$-process \cite{FranoschFuchsGoetze1997,FuchsGoetzeMayr1998}.  These higher order corrections cannot be read off directly from the simulation data, but have to be determined by fits.  In order to extract them the values calculated from the factorization theorem (first two terms in Eq.~(\ref{eq:quantbetaformula})) were subtracted from the simulation data.  Next, we chose an interval around $t_\mr{co}$ where fitting was performed.  We took $q$-independent intervals beginning and ending  approximately at the inflection points of the correlator before and after the plateau.  The dependence on the chosen interval was not very strong, but noticeable.  The short-time correction coefficients $B^\mr{x}_q$ are in general hard to determine as the microscopic dynamics influences the results.  The dependence of the early $\beta$-relaxation on the microscopic dynamics of the simulation was demonstrated e.g.\ in \cite{GleimKobBinder1998}. By comparing Newtonian and stochastic dynamics it was found that the different short-time behavior does not affect the late-$\beta$- and the $\alpha$-process, but the early $\beta$-process.  The MCT-prediction for the approach towards the plateau was visible for stochastic microscopic dynamics only.  Since the microscopic dynamics of our simulation is Newtonian, we cannot expect to obtain accurate results for $B^\mr{x}_q$ from fits of the simulation data.

\bfig
\rsfig{figures/short_vs_longtimecoeff+fit.eps}
\caption[]{
\label{fig:short_vs_longtimecoeff+fit}
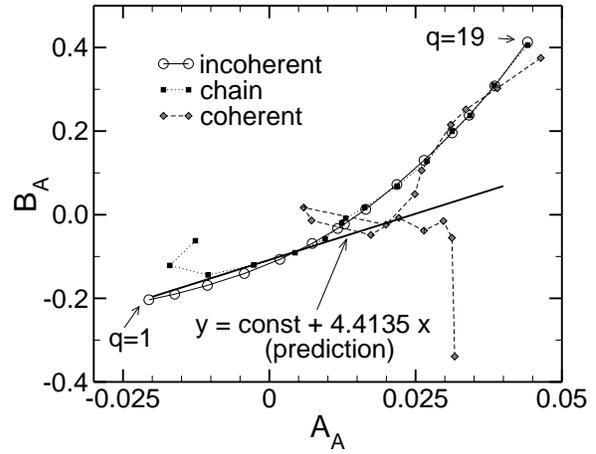
Short-time correction coefficients $B^\mr{x}_q$ as a function of the long-time correction coefficients $A^\mr{x}_q$ (shown in Fig.~\ref{fig:long_time_coeff_vs_q}) for $q = 1,2,3,4,5,6,6.9,7.15,8,9.5,11,12.8,14,16,19$.  MCT predicts a linear dependence with slope $1/B^2 = 4.4135$ (Eq.~(\ref{eq:B_A+A_A})), which is approximately satisfied at small $q$ for the incoherent scattering function.  For small $q$, extraction of $B_q$ from simulation data is very difficult, and  no smooth behaviour can be observed.  At large $q$, all curves $B_q^\mr{x}(A_q^\mr{x})$ roughly agree with one another, but here a precise determination of $A_q^\mr{x}$ is hampered by the low value of the $\beta$-plateau.
}
\efig

Figure~\ref{fig:long_time_coeff_vs_q} shows that the long-time correction parameters of the coherent melt scattering function $A_q$ are (approximately) in anti-phase relative to $S_q$.  This is in qualitative agreement with the findings for hard spheres \cite{FranoschFuchsGoetze1997} and diatomic molecules \cite{KaemmererKobSchilling1998}.  As in the case of the non-ergodicity parameters and the critical amplitudes, the values of $A^\mr{s}_q$ and $A^\mr{p}_q$ are very similar, not showing much structure.  

MCT predicts a linear relationship between $A^\mr{x}_q$ and $B^\mr{x}_q$ (cf.\ Eq.~(\ref{eq:B_A+A_A})).  Figure~\ref{fig:short_vs_longtimecoeff+fit} plots $B^\mr{x}_q$ versus $A^\mr{x}_q$ and compares the result with the theoretical prediction: a line with slope $1/B^2 = 4.4135$.  Whilst the curve for $B^\mr{s}_q(A^\mr{s}_q)$ exhibits linear behaviour with a slope close to $1/B^2$ as long as $q \le q_\mr{max}$, the curve bends upwards at larger $q$.  The coefficients of the coherent chain correlator behave similarly, except for very small $q$, where one observes deviations.  Contrary to that, $B_q(A_q)$ exhibits a very irregular structure.  There is no linear behaviour at all, presumably due to the problems mentioned above (oscillations at the beginning of the $\beta$-regime for $q < q_\mr{max}$, insufficient statistics) so that $B_q$ cannot be fitted reliably.  On the other hand, all curves almost coincide at large $q$, where the slope is significantly different from $1/B^2$.  In this $q$-range, however, one has to take into account that the small plateau value (cf.\ Figure~\ref{fig:NEPs+S_q_vs_q}) renders the determination of the long-time correction coefficient $A^\mr{x}_q$ difficult.  This might lead to deviations relative to the theoretical prediction.

Now, we want to test whether the higher order corrections improve the description of the $\beta$-process.  Since our focus is on the coherent scattering functions, we show comparisons of $\phi_q(t)$ and the MCT approximations in leading and next-to-lead\-ing order ($\phi_q^\mr{p}(t)$ is very similar to $\phi_q^\mr{s}(t)$ which has already been discussed in \cite{BennemannBaschnagelPaul1999_incoherent}).  In Figure~\ref{fig:1st+2nd_approx_coh_melt_sf_T047+T0.52_q_3_6.9_8_14} $\phi_q(t)$ is depicted for some $q$ and for the highest and lowest temperature for which a quantitative $\beta$-analysis was performed.  As pointed out above, the short-time correction coefficients $B_q$ could not be obtained reliably from the simulation data.  Therefore, we calculated them from the more reliable long-time correction coefficients $A_q$, using $B_q = -0.108 + 4.4135 A_q$, where the slope is $1/B^2$ and the offset is taken from the straight line in Figure~\ref{fig:short_vs_longtimecoeff+fit}.  For both temperatures, the factorization theorem provides a good description if one is not too far away from $t_\mr{co}$.  Especially in the late $\beta$-regime next-to-lead\-ing order corrections improve the description of the simulation data, contrary to the early $\beta$-regime, where the microscopic dynamics interfere and completely mask the critical decay.  Similar findings were reported in \cite{KaemmererKobSchilling1998} for diatomic molecules and in \cite{GleimKobBinder1998} for a binary LJ-mixture.

\bfig
\rsfig{figures/1st+2nd_approx_coh_melt_sf_T047_q_3_6.9_8_14.eps}
\rsfig{figures/1st+2nd_approx_coh_melt_sf_T052_q_3_6.9_8_14.eps}
\caption[]{
\label{fig:1st+2nd_approx_coh_melt_sf_T047+T0.52_q_3_6.9_8_14}
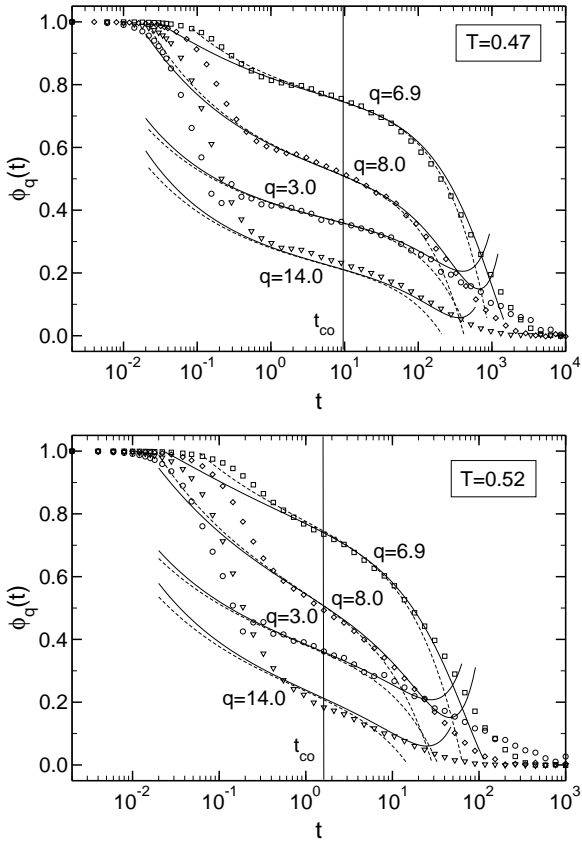
Coherent scattering function $\phi_q(t)$ from simulation data (symbols) for $q=3.0$, 6.9, 8.0, and 14.0 at $T=0.47$ (top) and $T=0.52$ (bottom), in comparison with MCT fits in the $\beta$-regime.  The vertical lines are at $t=t_\mr{co}$ ($\propto t_\sigma$), the centre of the $\beta$-regime.  Lowest order approximations (factorization theorem) are drawn with dashed lines, solid lines are used for the next-to-leading order approximations with short time correction coefficients computed from long-time correction coefficients (see Eq.~(\ref{eq:B_A+A_A})).  Note that different scales are used for the time axis in the upper and lower panels.  While the factorization theorem gives a good approximation around $t_\mr{co}$, higher corrections must be considered at later times.  As the fit results for $f^\mr{c}_q$ shows a significant temperature dependence at higher $q$ compared to small $q$, the temperature-averaged values for $f^\mr{c}_q$ used here lead to a visible offset of the curves for $q=14.0$.
}
\efig

\section{Conclusions}
\label{sec:conc}
In this paper we presented results from molecular-dy\-na\-mics simulations of a simple model for a supercooled non-en\-tang\-led polymer melt.  The temperatures investigated are within the range $T=0.46 \leq T \leq T=0.52$.  This range is above the critical temperature, $\Tc = 0.450 \pm 0.005$, of mode-coupling theory (MCT).  In order to obtain detailed information about the dynamics of the melt we computed the incoherent, the collective chain and the collective melt intermediate scattering functions as well as their space Fourier transforms, the van Hove correlation functions.  The main focus of this part of the paper was the relaxation of the melt in the $\beta$-regime.  The final structural $\alpha$-relaxation is the subject of the subsequent second part \cite{alphaDynamics}.   

It was found that the monomer displacements, $\vec{r}_i(t)-\vec{r}_j(t)$ ($i,j=1,\ldots,N$), of a chain are not Gaussian distributed at all times, as the Rouse model assumes, especially when temperature is close to $\Tc$ (Figs.~\ref{fig:gaussian_approx_q_1_2_3_4_5_69_95_14_19_T065.eps} and \ref{fig:gaussian_approx_q_1_2_3_4_5_69_95_14_T048.eps}).  However, the Rouse model provides a very good description of the mean-square displacements $\langle [\vec{r}_i(t) - \vec{r}_j(0)]^2 \rangle$.  This suggests that the Rouse modes of our model remain orthogonal at all times, even if the melt is strongly supercooled.  The same conclusion was also drawn in \cite{BennemannPaulBaschnagel1999_Rouse,GrantPaul_ChemPhys2000}.  This property could be important for the further advancement of theoretical models for supercooled polymer melts (see \cite{Guenza1999}, for instance).  

We applied tests, proposed in the literature \cite{Goetze1999_review,GleimKob2000,SignoriniBarratKlein1990,KobAndersen_LJ_I_1995,NaurothKob1997}, for the space-time factorization theorem of MCT, which work directly with the simulation data without invoking any fit procedure (Figs.~\ref{fig:R_A_T046} and \ref{fig:Hr_over_Hrp_all_vanhove_T048}).  These tests show that the theorem is verified in real and reciprocal space at all investigated temperatures and that the $q$-dependence of the next-to-lead\-ing order correction coefficients qualitatively agrees with MCT predictions.  Furthermore, the factorization property also seems to hold for the Rouse modes.  However, higher order corrections behave differently than those of the scattering functions (Fig.~\ref{fig:R_A_from_Rouse-modes_interpol+inc_sf_T048}).  By determining the critical amplitudes for the $\beta$-regime in real space conclusions about the typical length scales of the dynamics were obtained.  The $\beta$-pro\-cess emerges as a localized process which is dominated by the cooperative motion of a monomer and its nearest neighbours.  The influence of the other neighbour shells rapidly decreases to zero with increasing distance from the central monomer.

In a quantitative analysis all coefficients describing the correlators in the $\beta$-regime in leading and next-to-lead\-ing order in the separation parameter were computed and their $q$-dependence was discussed.  The analysis suggests that there are polymer-specific effects on the length scale of the coils, which contribute to the slowing down of structural relaxation on the scale of the chains (Figs.~\ref{fig:NEPs+S_q_vs_q} and \ref{fig:crit_ampl_vs_q}).  The  quantitative comparison with the MCT predictions corroborated our qualitative finding that the factorization theorem describes the simulated correlators well in the centre of the $\beta$-regime.  The parameters of the theorem, $f_q^\mr{x}$ and $h_q^\mr{x}$, for the coherent scattering functions were not fitted, but read off from the simulation data by applying general MCT predictions to the results obtained from a previous study of the incoherent scattering function \cite{BennemannBaschnagelPaul1999_incoherent}.  This shows that MCT can provide guidelines for a quantitative analysis of our simplified model of a supercooled polymer melt.  On the other hand, the next-to-lead\-ing order corrections have to be fitted.  They extend the quantitative description of the correlators significantly, especially at long times and at large wave-vectors (Fig.~\ref{fig:1st+2nd_approx_coh_melt_sf_T047+T0.52_q_3_6.9_8_14}), but a very precise determination of these correction terms over the whole $q$-range is hard to obtain by fits to the simulation data (Fig.~\ref{fig:short_vs_longtimecoeff+fit}).

\begin{acknowledgement}
We are indebted to C.\ Bennemann, J.\ Horbach, W.\ Kob, A.\ Latz, C.\ Brangian and F.\ Varnik for many helpful discussions and to M. Fuchs for valuable comments on the manuscript. This work would not have been possible without generous grants of computing time by the HLRZ J{\"u}\-lich, the RHRK Kai\-sers\-lau\-tern, the CTCMS at NIST, Gaithersburg, and the com\-put\-er centre at the University of Mainz. Financial support by the Deut\-sche For\-schungs\-ge\-mein\-schaft under grant No.\ SFB262/D2 and by the ESF Prog\-ramme on ``Experimental and Theoretical Investigation of Complex Polymer Structures'' (SUPERNET) is gratefully acknowledged.
\end{acknowledgement}

\bibliography{references_mct2}

\begin{thebibliography}{10}

\bibitem{Vigo1997}
K.~L. Ngai, E. Riande, and M.~D. Ingram~(eds.), J. Non-Cryst. Solids {\bf
  235-237},    (1998).

\bibitem{Pisa1998}
M. Giordano, D. Leporini, and M. Tosi~(eds.), J. Phys.: Condensed Matter {\bf
  11},  No.\ 10A  (1999).

\bibitem{Trieste1999}
S. Franz, S.~C. Glotzer, and S. Sastry~(eds.), J. Phys.: Condens. Matter {\bf
  12},  No.\ 29  (2000).

\bibitem{Goetze1999_review}
W. G{\"o}tze, J. Phys.: Condens. Matter {\bf 11},  A1  (1999).

\bibitem{goetzemctessentials}
W. G{\"o}tze, Condens. Mat. Phys. {\bf 1},  873  (1998).

\bibitem{GoetzeSjoegren1995_TTSP}
W. G\"otze and L. Sj\"ogren, Transport Theory Stat. Phys. {\bf 24},  801
  (1995).

\bibitem{Goetze_LesHouches}
W. G{\"o}tze,  in {\em Proceedings of the Les Houches Summer School of
  Theoretical Physics, Les Houches 1989, Session LI}, edited by J.~P. Hansen,
  D. Levesque, and J. Zinn-Justin (North-Holland, Amsterdam, 1991), pp.\
  287--503.

\bibitem{FuchsGoetzeHildebrand1992_extMCT}
M. Fuchs, W. G{\"o}tze, S. Hildebrand, and A. Latz, J. Phys.: Condens. Matter
  {\bf 4},  7709  (1992).

\bibitem{KirkpatrickThirumalai_TTSP1995}
T.~R. Kirkpatrick and D. Thirumalai, Transport Theory Stat. Phys. {\bf 24},
  927  (1995).

\bibitem{MezardParisi2000}
M. M\'ezard and G. Parisi, J. Phys.: Condens. Matter {\bf 12},  6655  (2000).

\bibitem{KrakoAlba2000}
V. Krakoviack and C. Alba-Simionesco, Europhys. Lett. {\bf 51},  420  (2000).

\bibitem{SchillingScheidsteger1997}
R. Schilling and T. Scheidsteger, Phys. Rev. E {\bf 56},  2932  (1997).

\bibitem{FabbianLatz2000}
L. Fabbian, A. Latz, R. Schilling, F. Sciortino, P. Tartaglia, and C. Theis,
  Phys. Rev. E {\bf 62},  2388  (2000).

\bibitem{FranoschGoetze_orient1997}
T. Franosch, W. G\"otze, M. Fuchs, M.~R. Mayr, and A.~P. Singh, Phys. Rev. E
  {\bf 56},  5659  (1997).

\bibitem{FranoschGoetze1997}
T. Franosch, W. G\"otze, M.~R. Mayr, and A.~P. Singh, Phys. Rev. E {\bf 55},
  3183  (1997).

\bibitem{goetzevoigtmann2000}
W. G\"otze and T. Voigtmann, Phys. Rev. E {\bf 61},  4133  (2000).

\bibitem{Latz2000}
A. Latz, J. Phys.: Condens. Matter {\bf 12},  6353  (2000).

\bibitem{kobreview1999}
W. Kob, J. Phys.: Condens. Matter {\bf 11},  R85  (1999).

\bibitem{DonatiGlotzer1999}
C. Donati, S.~C. Glotzer, P.~H. Poole, W. Kob, and S.~J. Plimpton, Phys. Rev. E
  {\bf 60},  3107  (1999).

\bibitem{Allegrini1999}
P. Allegrini, J.~F. Douglas, and S.~C. Glotzer, Phys. Rev. E {\bf 60},  5714
  (1999).

\bibitem{YamamotoOnuki1998}
R. Yamamoto and A. Onuki, Phys. Rev. E {\bf 58},  3515  (1998).

\bibitem{DoliwaHeuer1998}
B. Doliwa and A. Heuer, Phys. Rev. Lett. {\bf 80},  4915  (1998).

\bibitem{DoliwaHeuer2000}
B. Doliwa and A. Heuer, Phys. Rev. E {\bf 61},  6898  (2000).

\bibitem{VollmayrKob2000}
K. Vollmayr-Lee, W. Kob, K. Binder, and A. Zippelius, Int. J. Mod. Phys. C {\bf
  10},  1443  (2000).

\bibitem{BuechnerHeuer_PRE1999}
S. B\"uchner and A. Heuer, Phys. Rev. E {\bf 60},  6507  (1999).

\bibitem{BuechnerHeuer_PRL2000}
S. B\"uchner and A. Heuer, Phys. Rev. Lett. {\bf 84},  2168  (2000).

\bibitem{SastryDebenedetti1999}
S. Sastry, P.~G. Debenedetti, F.~H. Stillinger, J.~C. Schr{\o}der, T. B.~Dyre,
  and S.~C. Glotzer, Physica A {\bf 270},  301  (1999).

\bibitem{SchroderSastry2000}
T.~B. Schr{\o}der, S. Sastry, J.~C. Dyre, and S.~C. Glotzer, J. Chem. Phys.
  {\bf 112},  9834  (2000).

\bibitem{SciortinoKobTartaglia2000}
F. Sciortino, W. Kob, and P. Tartaglia, J. Phys.: Condens. Matter {\bf 12},
  6525  (2000).

\bibitem{KobBarrat2000}
W. Kob and J.-L. Barrat, Eur. Phys. J. B {\bf 13},  319  (2000).

\bibitem{KobSciortinoTartaglia2000}
W. Kob, F. Sciortino, and P. Tartaglia, Europhys. Lett. {\bf 49},  590  (2000).

\bibitem{KaemmererKobSchilling1998}
S. K{\"a}mmerer, W. Kob, and R. Schilling, Phys. Rev. E {\bf 58},  2131
  (1998).

\bibitem{KaemmererKobSchilling1998_orient}
S. K{\"a}mmerer, W. Kob, and R. Schilling, Phys. Rev. E {\bf 58},  2141
  (1998).

\bibitem{KaemmererKobSchilling1997}
S. K{\"a}mmerer, W. Kob, and R. Schilling, Phys. Rev. E {\bf 56},  5450
  (1997).

\bibitem{TheisSciortino2000}
C. Theis, F. Sciortino, A. Latz, R. Schilling, and P. Tartaglia, Phys. Rev. E
  {\bf 62},  1856  (2000).

\bibitem{LewisWahnstrom1994}
L.~J. Lewis and G. Wahnstr\"om, Phys. Rev. E {\bf 50},  3865  (1994).

\bibitem{Mossa2000}
S. Mossa, R. Di~Leonardo, G. Ruocco, and M. Sampoli, Phys. Rev. E {\bf 62},
  612  (2000).

\bibitem{sciortino1996}
F. Sciortino, P. Gallo, P. Tartaglia, and S.-H. Chen, Phys. Rev. E {\bf 54},
  6331  (1996).

\bibitem{SciortinoFabbianChen1997}
F. Sciortino, L. Fabbian, S.-H. Chen, and P. Tartaglia, Phys. Rev. A {\bf 56},
  5397  (1997).

\bibitem{Sciortino2000}
F. Sciortino, Chem. Phys. {\bf 258},  307  (2000).

\bibitem{StarrSciortino1999}
F.~W. Starr, F. Sciortino, and H.~E. Stanley, Phys. Rev. E {\bf 60},  6757
  (1999).

\bibitem{ZonLeeuw1998}
A. van Zon and S.~W. de~Leeuw, Phys. Rev. E {\bf 58},  R4100  (1998).

\bibitem{ZonLeeuw1999}
A. van Zon and S.~W. de~Leeuw, Phys. Rev. E {\bf 60},  6942  (1999).

\bibitem{BBPB_2000}
J. Baschnagel, C. Bennemann, W. Paul, and K. Binder, J. Phys.: Condens. Matter
  {\bf 12},  6365  (2000).

\bibitem{alphaDynamics}
M. Aichele and J. Baschnagel, Eur. Phys. J. E {\bf xx},  000  (2001), to be
  filled in.

\bibitem{BennemannBaschnagelPaul1999_incoherent}
C. Bennemann, J. Baschnagel, and W. Paul, Eur. Phys. J. B {\bf 10},  323
  (1999).

\bibitem{BennemannPaulBinder1998}
C. Bennemann, W. Paul, K. Binder, and B. D{\"u}nweg, Phys. Rev. E {\bf 57},
  843  (1998).

\bibitem{BennemannPaulBaschnagel1999}
C. Bennemann, W. Paul, J. Baschnagel, and K. Binder, J. Phys.: Condens. Matter
  {\bf 11},  2179  (1999).

\bibitem{BennemannPaulBaschnagel1999_Rouse}
C. Bennemann, J. Baschnagel, W. Paul, and K. Binder, Comp. Theo. Poly. Sci.
  {\bf 9},  217  (1999).

\bibitem{natureBDBG1999}
C. Bennemann, C. Donati, J. Baschnagel, and S.~C. Glotzer, Nature {\bf 399},
  246  (1999).

\bibitem{Aichele_DH2000}
M. Aichele, J. Baschnagel, and S.~C. Glotzer, in preparation  (2001).

\bibitem{KremerGrest1990}
K. Kremer and G.~S. Grest, J. Chem. Phys. {\bf 92},  5057  (1990).

\bibitem{MeyerMuellerPlate2001}
H. Meyer and F. M\"uller-Plate, preprint  (2001).

\bibitem{DoiEdwards}
M. Doi and S.~F. Edwards, {\em The Theory of Polymer Dynamics} (Oxford
  University Press, Oxford, 1986).

\bibitem{HansenMcDonald}
J.~P. Hansen and I.~R. McDonald, {\em Theory of Simple Liquids} (Academic
  Press, London, 1986).

\bibitem{FranoschFuchsGoetze1997}
T. Franosch, M. Fuchs, W. G{\"o}tze, M.~R. Mayr, and A.~P. Singh, Phys. Rev. E
  {\bf 55},  7153  (1997).

\bibitem{FuchsGoetzeMayr1998}
M. Fuchs, W. G{\"o}tze, and M.~R. Mayr, Phys. Rev. E {\bf 58},  3384  (1998).

\bibitem{KremerGrest_review1995}
K. Kremer and G.~S. Grest,  in {\em Monte Carlo and Molecular Dynamics
  Simulations in Polymer Science}, edited by K. Binder (Oxford University
  Press, New York, 1995), pp.\ 194--271.

\bibitem{BinderPaul_review1997}
K. Binder and W. Paul, J. Polym. Sci. B {\bf 35},  1  (1997).

\bibitem{PuetzKremerGrest2000}
M. P\"utz, K. Kremer, and G.~S. Grest, Europhys. Lett. {\bf 49},  735  (2000).

\bibitem{Verdier1966}
P.~H. Verdier, J. Chem. Phys. {\bf 45},  2118  (1966).

\bibitem{paul1998}
W. Paul, G.~D. Smith, D.~Y. Yoon, B. Farago, S. Rathgeber, A. Zirkel, L.
  Willner, and D. Richter, Phys. Rev. Lett. {\bf 80},  2346  (1998).

\bibitem{GrantPaul2000}
G.~D. Smith, W. Paul, M. Monkenbusch, and D. Richter, J. Chem. Phys.  (in
  press).

\bibitem{GrantPaul_ChemPhys2000}
G.~D. Smith, W. Paul, M. Monkenbusch, and D. Richter, Chem. Phys. {\bf 261},
  61  (2000).

\bibitem{GleimKob2000}
T. Gleim and W. Kob, Eur. Phys. J. B {\bf 13},  83  (2000).

\bibitem{SignoriniBarratKlein1990}
G.~F. Signorini, J.-L. Barrat, and M.~L. Klein, J. Chem. Phys. {\bf 92},  1294
  (1990).

\bibitem{KobAndersen_LJ_I_1995}
W. Kob and H.~C. Andersen, Phys. Rev. E {\bf 51},  4626  (1995).

\bibitem{ToelleSchoberWuttke1997}
A. T\"olle, H. Schober, J. Wuttke, and F. Fujara, Phys. Rev. E {\bf 56},  809
  (1997).

\bibitem{WuttkeSeidl1998}
J. Wuttke, M. Seidl, G. Hinze, A. T\"olle, and G. Coddens, Eur. Phys. J. B {\bf
  1},  169  (1998).

\bibitem{LunkenheimerReview2000}
P. Lunkenheimer, U. Schneider, R. Brand, and A. Loidl, Contempory Phys. {\bf
  41},  15  (2000).

\bibitem{NaurothKob1997}
M. Nauroth and W. Kob, Phys. Rev. E {\bf 55},  657  (1997).

\bibitem{BarratGoetzeLatz1989}
J.-L. Barrat, W. G{\"o}tze, and A. Latz, J. Phys.: Condens. Matter {\bf 1},
  7163  (1989).

\bibitem{KobAndersen_LJ_II_1995}
W. Kob and H.~C. Andersen, Phys. Rev. E {\bf 52},  4134  (1995).

\bibitem{GleimKobBinder1998}
T. Gleim, W. Kob, and K. Binder, Phys. Rev. Lett. {\bf 81},  4404  (1998).

\bibitem{Guenza1999}
M. Guenza, J. Chem. Phys. {\bf 110},  7574  (1999).

\end{thebibliography}

\end{document}